\documentclass[a4paper,11pt]{article}
\usepackage{jheppub} 

\usepackage{bm}
\usepackage{color}
\usepackage{amssymb}
\usepackage{amsmath}
\usepackage{subfigure}

\usepackage{MnSymbol}
\usepackage{multirow}
\usepackage{diagbox}

\newcommand{\blue}[1]{{\color{blue}  #1}}

\newcommand{\bs}[1]{\boldsymbol{#1}}

\begin{document}

\title{The connection between Nucleon Energy Correlators and Fracture Functions}

\author[a]{Kai-Bao Chen}
\emailAdd{chenkaibao19@sdjzu.edu.cn}

\author[b,c,d]{Jian-Ping Ma}
\emailAdd{majp@itp.ac.cn}

\author[e,f]{Xuan-Bo Tong}\emailAdd{xuan.bo.tong@jyu.fi} 

\affiliation[a]{School of Science, Shandong Jianzhu University, Jinan, Shandong 250101, China}
\affiliation[b]{CAS Key Laboratory of Theoretical Physics, Institute of Theoretical Physics, P.O. Box 2735, Chinese Academy of Sciences, Beijing 100190, China}
\affiliation[c]{School of Physical Sciences, University of Chinese Academy of Sciences, Beijing 100049, China}
\affiliation[d]{School of Physics and Center for High-Energy Physics, Peking University, Beijing 100871, China}
\affiliation[e]{Department of Physics, University of Jyväskylä, P.O. Box 35, 40014 University of Jyväskylä, Finland}
\affiliation[f]{Helsinki Institute of Physics, P.O. Box 64, 00014 University of Helsinki, Finland}
\abstract{We establish a sum rule that connects fracture functions to nucleon energy-energy correlators~(NEECs) in a one-to-one correspondence. Using this sum rule, we study the energy pattern in the target fragmentation region of deep inelastic scatterings. Through investigations up to twist-3, we express all eighteen energy-pattern structure functions in terms of associated NEECs, elucidating various azimuthal and spin asymmetries critical for nucleon tomography. Additionally, we investigate the perturbative matching of the twist-2 quark NEECs. We demonstrate that the Sivers-type and worm-gear-type quark NEECs match onto twist-3 multi-parton distributions. Our work provides a framework for examining energy-weighted observables through hadron production processes in the target fragmentation region, offering new insights into nucleon tomography.

}

\maketitle
\flushbottom

\newpage
\section{Introduction}
\label{sec:Introduction}
Nucleon tomography has been a central focus in hadron physics over recent decades, playing a key role in experiments at facilities such as HERA~\cite{Klein:2008di}, JLab~\cite{Burkert:2020akg,Accardi:2023chb}, and upcoming electron-ion colliders~(EIC)~\cite{Boer:2011fh,Accardi:2012qut,AbdulKhalek:2021gbh,Anderle:2021wcy}. An important process in this field is Semi-Inclusive Deep Inelastic Scattering (SIDIS), which offers insights into nucleon structure by detecting an additional hadron $h$ in the DIS. Significant progress in understanding parton distribution functions (PDFs) and fragmentation functions (FFs) has been made through analyses of hadron production in the current fragmentation region (CFR). These advances have been facilitated by both transverse-momentum-dependent~(TMD)~\cite{Ji:2004wu,Ji:2004xq,Collins:2011zzd,Boussarie:2023izj} and collinear factorization frameworks~(see e.g.,~\cite{Bacchetta:2008xw,Benic:2019zvg,Benic:2021gya,Benic:2024fvk,Goyal:2023xfi,Bonino:2024qbh} and the references therein), as well as the small-$x$ formalism~\cite{Marquet:2009ca,Iancu:2020jch,Tong:2022zwp,Bergabo:2022zhe,Bergabo:2024ivx,Caucal:2024cdq}. A common feature of these tomographic studies is the ability to distinguish between the parton dynamics inside the initial target and those responsible for fragmentation into the hadron $h$.

However, such separation is generally not feasible for SIDIS in the target fragmentation region (TFR), where the detected hadron $h$ moves into the forward region of the incoming nucleon. This challenge was first addressed by Trentadue and Veneziano~\cite{Trentadue:1993ka}, who introduced fracture functions to capture the complex interplay between the initial and final-state dynamics in the TFR. \emph{Fracture functions}%
\footnote{It is noted that in the original definition of fracture functions proposed in~\cite{Trentadue:1993ka}, the dependence on the hadron transverse momentum $\bs P_{h\perp}$ was integrated out. However, it was soon realized in~\cite{Berera:1995fj,Grazzini:1997ih,Collins:1997sr} that this integration is not necessary. They extended the definition to incorporate the $\bs P_{h\perp}$-dependence, which is now commonly used in the studies of fracture functions~(e.g.,~\cite{Anselmino:2011ss,Chen:2021vby,Chen:2024brp,Chen:2023wsi,Guo:2023uis}). We will focus on this extended version. In addition, it is useful to recall that the term ``\emph{fracture}'' was conied to describe the partonic \emph{structure} of the target once it \emph{fragments} into a given hadron $h$.}%
%
~specifically describe the distributions of the struck parton inside the target when the spectator partons fragment into a specific hadron $h$~\cite{Trentadue:1993ka,Berera:1995fj,Grazzini:1997ih}. The universality of these functions for deep-inelastic processes has been demonstrated in seminal studies~\cite{Collins:1997sr,Grazzini:1997ih}, and their effectiveness in describing nucleon structure in the TFR has been shown in various phenomenological analyses~\cite{deFlorian:1998rj,Goharipour:2018yov,Khanpour:2019pzq,Maktoubian:2019ppi,Salajegheh:2022vyv,Salajegheh:2023jgi,deFlorian:1997wi,Ceccopieri:2014rpa,Shoeibi:2017lrl,Shoeibi:2017zha,Ceccopieri:2012rm,Ceccopieri:2015kya}. Additionally, recent works have achieved complete one-loop calculations for spin and azimuthal-dependent structure functions~(SFs) in the TFR SIDIS, highlighting the critical role of gluonic fracture functions in generating azimuthal asymmetries~\cite{Chen:2024brp}. Further, tree-level twist-3 contributions to SIDIS in the TFR have been derived in~\cite{Chen:2023wsi}. All eighteen SIDIS structure functions in the TFR are now predicted in terms of associated fracture functions. Moreover, the perturbative matching of Sivers- and worm-gear-type fracture functions onto twist-3 multi-parton correlation functions~\cite{Efremov:1981sh,Efremov:1983eb,Qiu:1991pp,Qiu:1998ia,Ji:1992eu,Eguchi:2006qz,Eguchi:2006mc} has been studied in~\cite{Chen:2021vby}, accounting for the transition of single-spin asymmetry~(SSA) and double-spin asymmetry~(DSA) between the TFR and the CFR.  

In fact, by selecting different quantum numbers of the detected hadron $h$, one can resolve various colored and flavored contents inside the target~(see e.g.,~\cite{Ceccopieri:2012rm,Ceccopieri:2014rpa,Goharipour:2018yov}). For instance, when $h$ is a diffractive nucleon, the associated fracture functions can encode pomeron exchanges and are often termed as diffractive PDFs~\cite{Berera:1995fj,Collins:1997sr}.  Moreover, TMD fracture functions, which incorporate partonic transverse momentum dependence, have been introduced~\cite{Anselmino:2011ss,Chai:2019ykk,Hatta:2022lzj} with several observables proposed to measure these functions~\cite{Anselmino:2011ss,Hatta:2022lzj,Anselmino:2011vkz,Anselmino:2011bb,Chai:2019ykk,CLAS:2022sqt,Guo:2023uis,Iancu:2021rup,Iancu:2022lcw,Hauksson:2024bvv,Tong:2023bus,Shao:2024nor}. Their evolution equations are studied in~\cite{Chai:2019ykk,Hauksson:2024bvv}. Particularly, diffractive TMD fracture functions have recently garnered attention for their potential in revealing small-$x$ dynamics~\cite{Hatta:2022lzj,Hatta:2024vzv,Iancu:2021rup,Iancu:2022lcw,Hauksson:2024bvv,Tong:2023bus,Shao:2024nor}.

 \begin{figure}[htpp]
    \centering
    \includegraphics[scale=0.65]{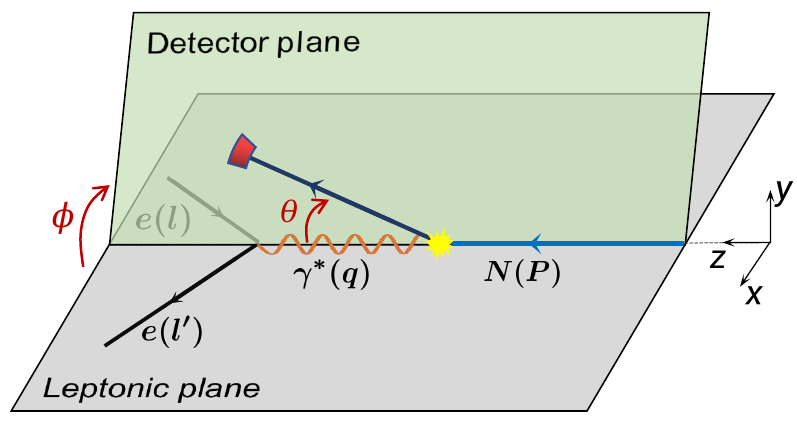}
    \caption{Illustration of the DIS energy pattern in the photon-nucleon collinear frame. }
    \label{fig:kinematics}
\end{figure}

In addition to SIDIS, the \emph{Energy-Energy Correlator} (EEC) has recently emerged as a novel tool to study nucleon tomography within DIS in both the CFR~\cite{Li:2021txc,Kang:2023big} and TFR~\cite{Liu:2022wop,Cao:2023oef,Liu:2023aqb,Li:2023gkh,Liu:2024kqt,Cao:2023qat,Guo:2024jch}. As an event-shape observable, the EEC captures the angular correlations among the asymptotic hadronic energy flows in reactions. It was originally proposed in $e^+e^-$ collisions~\cite{Basham:1977iq,Basham:1978bw,Basham:1978zq} as a precise test of perturbative quantum chromodynamics~(QCD). Recent years have seen significant progress in the studies of EEC~\cite{Tulipant:2017ybb,Dixon:2018qgp,Luo:2019nig,DelDuca:2016ily,Henn:2019gkr,Moult:2018jzp, Ebert:2020sfi,Duhr:2022yyp, Moult:2019vou,Dixon:2019uzg,Korchemsky:2019nzm,Chen:2023wah,Yang:2024gcn,Chen:2022pdu,Chen:2020adz,Chen:2022swd,Craft:2022kdo,Komiske:2022enw,Jaarsma:2023ell,Lee:2022ige,Lee:2023xzv,Lee:2023npz,Lee:2023tkr,CMS:2024mlf,Andres:2022ovj,Andres:2023xwr,Devereaux:2023vjz,Andres:2023ymw,Yang:2023dwc,Barata:2023bhh,Barata:2023zqg,Chen:2024nfl}, including investigations into jet substructures~\cite{Chen:2022pdu,Chen:2020adz,Chen:2022swd,Craft:2022kdo,Komiske:2022enw,Jaarsma:2023ell,Lee:2022ige,Lee:2023xzv,Lee:2023npz,Lee:2023tkr,CMS:2024mlf} and QCD medium effects~\cite{Andres:2022ovj,Andres:2023xwr,Devereaux:2023vjz,Andres:2023ymw,Yang:2023dwc,Barata:2023bhh,Barata:2023zqg} in hadron colliders. For DIS, an intriguing adaptation under study is to measure the angular distribution of a single hadronic energy flow in the photon-nucleon collinear frame~\cite{Li:2021txc,Kang:2023big,Liu:2022wop,Cao:2023oef,Liu:2023aqb,Li:2023gkh,Liu:2024kqt}. This observable was first proposed in \cite{Meng:1991da} as an extension of the energy pattern cross section studied in $e^+e^-$ collisions~\cite{Basham:1977iq}. Unlike SIDIS requiring hadron identification, the \emph{DIS energy pattern}
 \footnote{The DIS energy pattern introduced here corresponds to the same observable referred to as ``\emph{EEC in DIS}'' in ~\cite{Li:2021txc,Kang:2023big} and ``\textit{energy-weighted cross section}'' in ~\cite{Liu:2023aqb,Liu:2024kqt}. Its Mellin moment is referred to as ``\emph{$x_B$-weighted EEC}" in \cite{Liu:2022wop,Cao:2023oef,Cao:2023qat}. We follow~\cite{Basham:1977iq} and use the term,``\emph{energy pattern}'', specifically to avoid the ambiguity with the \emph{nucleon EEC}, which is a parton correlation matrix used to factorize the energy pattern cross section in the TFR. Moreover, the term ``\emph{energy pattern}'' specifically denotes the distribution of a single energy flow, which can be interpreted as an antenna pattern in reactions~\cite{Basham:1977iq}.}
can be readily measured by recording the energy deposits in the calorimeter at specific solid angles $\Omega=(\theta,\phi)$. Here, $\theta$ represents the polar angle of the calorimeter with the nucleon beam direction as the $z$-axis, and $\phi$ denotes the azimuthal angle relative to the lepton plane. This configuration is depicted in figure~\ref{fig:kinematics}.

 Despite the differences in experimental measurements, the DIS energy pattern cross section can be effectively computed from the differential cross section of SIDIS~(e.g.,~\cite{Li:2021txc,Kang:2023big,Liu:2023aqb}):
 \begin{align}
 \Sigma(\theta,\phi) =&\sum_h \int d \sigma^{e +N\rightarrow e+h+X} \frac{E_h}{E_N}\delta(\theta^2-\theta_h^2) \delta(\phi-\phi_h)~.
 \label{eq:energy pattern}
\end{align}
Here, the phase-space integral on the right-hand side pretains to the production of a hadron $h$ into given angles $(\theta,\phi)$, modified by the insertion of an energy weighting factor $E_h/E_N$ in the integrand, where $E_N$ and $E_h$ represent the energies of the nucleon beam and the hadron $h$, respectively. The summation is carried out over all possible hadron types $h$. Following SIDIS, the DIS energy pattern contains a total of eighteen SFs, when incorporating polarization effects of the nucleon target and the lepton beam~\cite{Meng:1991da}. Furthermore, the energy deposit in the region $\pi-\theta\ll 1$ comes from the hadrons in the CFR, while that in the region $\theta\ll1$ is from the TFR.

Applying the aforementioned relationship in the CFR, refs.~\cite{Li:2021txc,Kang:2023big} utilized the TMD studies of SIDIS and demonstrated that the DIS energy pattern provides new probes into the conventional TMD PDFs. The initial application, where the TMD factorization for the azimuthal- and spin-averaged energy pattern was derived, was presented in ref.~\cite{Li:2021txc}. Subsequently, ref.~\cite{Kang:2023big} extended the study to include the spin and azimuthal dependencies, contributing to seven additional EPSFs and elucidating mechanisms of the Sivers- and Collins-type asymmetries. Furthermore, a similar approach has also been employed in exploring the transverse energy correlator in DIS~\cite{Li:2020bub,Kang:2023oqj}.

However, the studies of the DIS energy pattern in the TFR~\cite{Liu:2022wop,Cao:2023oef,Liu:2023aqb,Li:2023gkh,Liu:2024kqt} have been conducted entirely independently from those on SIDIS. Notably, instead of fracture functions, a novel quantity known as the \emph{Nucleon Energy-Energy Correlator} (NEEC)~\cite{Liu:2022wop} has been introduced to describe the factorization of the energy pattern in the TFR~\cite{Cao:2023oef}. This newly introduced correlator characterizes the correlations between the initiation of a parton with momentum fraction $x$ from the target and the formation of an energy flow at a given angle $(\theta,\phi)$ in the TFR. It has been argued that the NEEC can encode Sivers effects and induce a SSA~\cite{Liu:2022wop,Liu:2024kqt}. Moreover, it accommodates the presence of linearly polarized gluons, which manifest as a $\cos 2\phi$-asymmetry~\cite{Li:2021txc,Guo:2024jch}. Additionally, the NEEC provides probes into small-$x$ gluons and saturation effects~\cite{Liu:2023aqb}. It is also shown that the information of the TMD PDFs can be extracted from the NEECs and their extensions, semi-inclusive energy correlators~\cite{Liu:2024kqt}.

The aim of this paper is to bridge the gap between the emerging investigations of the DIS energy pattern and the recent advancements of SIDIS in the TFR. In particular, we establish a sum rule that connects the conventional fracture functions to the novel NEECs. This sum rule serves as a parton-level proxy of the cross-section connection in eq.~(\ref{eq:energy pattern}), revealing that fracture functions can be regarded as the parent functions of NEECs. Importantly, this relationship allows the NEECs to inherit the essential initial- and final-state correlations originally encoded in fracture functions. Consequently, we find that fracture functions and NEECs exhibit a one-to-one correspondence. For instance, the NEEC that incorporates Sivers effects~\cite{Liu:2022wop,Liu:2024kqt} is due to the non-zero Sivers-type fracture function previously studied in~\cite{Anselmino:2011ss,Chen:2021vby}. Similarly, the counterpart of the linearly polarized gluon NEEC~\cite{Li:2023gkh} can be found in a recent work~\cite{Chen:2024brp}.  Furthermore, this sum rule suggests that additional novel NEECs can be identified from earlier studies of fracture functions, and it remains valid beyond the leading twist level. Additionally, we also present the sum rules for nucleon $N$-point EECs and semi-inclusive energy correlators.

As an application of this sum rule, we conduct a thorough investigation on the DIS energy pattern in the TFR. We note that while the factorization in terms of NEECs was studied in \cite{Liu:2022wop,Cao:2023oef}, these studies primarily focus on the leading-twist contributions to the azimuthal- and spin-independent SFs. In this work, we delve into all the eighteen energy-pattern SFs, expressing each SF in terms of associated NEECs and including contributions beyond leading twist. These results are derived from our recent studies of the TFR SIDIS in~\cite{Chen:2023wsi,Chen:2024brp}. Specifically, we find that ten SFs are contributed by the twist-2 NEECs, with four exclusively involving gluonic NEECs. Additionally, the remaining eight SFs are generated by twist-3 NEECs. Furthermore, we derive contributions to various azimuthal and spin asymmetries, providing new avenues for nucleon tomography.

We also investigate the perturbative matching of the twist-2 quark NEECs through the sum rule. It is shown that the NEECs at moderately large $\theta$ can be matched onto the collinear parton distributions, providing insight into the transition between the TFR and CFR~\cite{Liu:2022wop,Cao:2023oef,Cao:2023qat}. Resummation based on the perturbative matching has been also conducted~\cite{Cao:2023oef,Cao:2023qat}. However, the investigations so far have only focused on the $n_t$-even quark NEECs, namely those survive after the $\phi$ integration. Here, $n_t$ refers to the azimuthal vector of the energy flow. These NEECs only receive leading-twist two-parton correlations in the matching calculation. Based on our previous matching studies of fracture functions in~\cite{Chen:2021vby}, we perform the first study on the $n_t$-odd NEECs, specifically the Sivers-type quark NEEC and the worm-gear-type quark NEEC. The matching of these two NEECs is non-trivial, as they necessitate twist-3 contributions and involve multi-parton correlations. In particular, while the worm-gear NEEC accounts for T-even effects and induces a DSA, the Sivers NEEC generates a SSA, which is a T-odd effect and requires a nontrivial phase to be generated in the perturbative region.

The rest of this paper is organized as follows.
In section~\ref{sec:connection}, we establish the sum rules between NEECs and fracture functions. In section~\ref{sec:SF}, we present the classifications of the energy patten SFs and illustrate their relations to SIDIS SFs.  In section~\ref{sec:SFfactorization}, we derive the factorization of the the energy pattern SFs in the TFR in terms of NEECs. In section~\ref{sec:matching}, we present the perturbative matching for the quark NEECs.
Section~\ref{sec:summary} provides a summary of our findings and conclusions.

\section{NEECs and Fracture functions\label{sec:connection}}   

 Through this manuscript, we use the light-cone variables, in which a four-vector $a^\mu$ is expressed as $a^\mu = (a^+,a^-, \bs a_\perp) = \bigl((a^0+a^3)/{\sqrt{2}}, (a^0-a^3)/{\sqrt{2}}, a^1, a^2 \bigr)$. We introduce the light-cone vectors $n^\mu = (0,1,0,0)$ and $\bar n^\mu = (1,0,0,0)$, and define the transverse metric as $g_\perp^{\mu\nu} = g^{\mu\nu} - \bar n^\mu n^\nu - \bar n^\nu n^\mu$. The transverse antisymmetric tensor is given as $\varepsilon_\perp^{\mu\nu} = \varepsilon^{\mu\nu\alpha\beta} \bar n_\alpha n_\beta$ with $\varepsilon^{0123}=1$ and $\varepsilon_\perp^{12} = 1$. We also introduce the notation $\tilde a_\perp^\mu \equiv \varepsilon_\perp^{\mu\nu} a_{\perp\nu}$ for convenience.

\subsection{The connection: quark sector \label{sec:quarkNEEC}}
\subsubsection{Sum rules for the correlation matrices}
Let us start by establishing the sum rule for the connection between fracture functions and NEECs. We focus on collinear quark contributions from a nucleon target in this subsection. The gluonic case will be addressed later in subsection \ref{sec:gluonNEECs}, with additional extensions discussed in appendices \ref{sec:semiEEC} and \ref{sec:TMD_NEEC}. We assume that the nucleon target moves rapidly in the positive $z$-direction, characterized by the momentum $P^\mu=(P^+,P^-,\bs 0_\perp)$ satisfying $P^+\gg P^-$, and is polarized with the vector $S^\mu$:
\begin{align}
S^\mu = S_L \frac{P^+}{M} \bar n^\mu - S_L \frac{M}{2P^+} n^\mu+ S_\perp^\mu~,
\label{eq:spinvector}
\end{align}
where $S_L$ denotes the nucleon helicity, and $S_\perp^\mu$ represents the nucleon transverse polarization vector. $M$ is the nucleon mass. 

In the context of fracture functions, we measure a hadron $h$ in the forward region of the target nucleon. It is convenient to introduce the longitudinal momentum fraction $\xi_h=P_h^+/P^+$ and parameterize the momentum as 
\begin{align}
P_h^\mu=\Big(\xi_h P^+, \frac{\bs P_{h\perp}^2}{2\xi_h P^+}, \bs P_{h\perp}\Big)~.
\end{align}
The associated quark fracture functions are then defined through the following correlation matrix~(see e.g.,~\cite{Chen:2023wsi}): 
\begin{align}
{\cal M}^{q}_{ij,\text{FrF}}(x,\xi_h,\bs P_{h\perp}) =&  \int \frac{d\eta^-}{2\xi_h(2\pi)^4} e^{-ixP^+\eta^-} \sum_X\int\frac{d^3 \bs P_X}{2 E_X(2\pi)^3}
\notag \\ 
& \times \langle PS|\bar \psi_j(\eta^-) {\cal L}_n^{\dagger}(\eta^-) |P_h X \rangle \langle X P_h| {\cal L}_n(0) \psi_i(0) |PS\rangle~,
\label{eq:M2p}
\end{align} 
where $i,j$ denote Dirac and color indices from the quark field $\psi$, and the light-cone gauge link is defined as ${\cal L}_n (x) = \text{P} \exp [ - i g_s \int_0^\infty d\lambda ~A^{+}(\lambda n +x) ]~$ with $A^\mu=A^{a,\mu}t^a$ as the gluon field in the fundamental representation. The sum $\sum_X$ represents the summation over all the unidentified out states $X$. We also sum over the polarization of the detected hadron $h$, if present.

Rather than focusing on a specific type of hadron, NEECs capture the energy flow from all possible hadrons in a given direction within the TFR. The quark contributions to the NEECs are defined from the following correlation matrix~\cite{Liu:2022wop}:
\begin{align}
{\cal M}_{ij,{\rm EEC}}^{q}(x,\theta,\phi) =& \int \frac{d\eta^-}{2\pi} e^{-ix P^+ \eta^-}\sum_X\int\frac{d^3 \bs P_X}{2 E_X(2\pi)^3}\sum_{a \in X} \delta(\theta^2-\theta^2_a)\delta(\phi-\phi_a)\frac{E_a}{E_N}
\notag \\ &\times  \langle PS|\bar \psi_j(\eta^-) {\cal L}_n^{\dagger}(\eta^-)|X \rangle \langle X|
{\cal L}_n(0) \psi_i(0) |PS\rangle~. \label{eq:NEEC}
\end{align}
Here, the sum $\sum_X$ spans a complete set of out states $X$, and $\sum_{a \in X}$ iterates over all particles within a given state $X$. Each particle's contribution is weighted by its energy $E_a$ normalized to the target energy $E_N$. The delta functions kinematically restrict the particles to those forming the energy flow in the specified solid angle $(\theta,\phi)$. The polar angle $\theta$ is measured relative to the target beam direction, and the azimuthal $\phi$-plane is perpendicular to this direction. 

Utilizing the normalized energy flow operator~\cite{Sveshnikov:1995vi,Korchemsky:1997sy,Tkachov:1995kk,Korchemsky:1999kt,Bauer:2008dt,Cao:2023oef}, defined as 
\begin{align}
{\cal E}(\theta,\phi)|X \rangle=\sum_{a \in X} \delta(\theta^2-\theta^2_a)\delta(\phi-\phi_a)\frac{E_a}{E_N}|X \rangle~,
\label{eq:Eflow}
\end{align}
the correlation matrix can be compactly rewritten as~\cite{Liu:2022wop,Liu:2024kqt}:
\begin{align}
{\cal M}_{ij,{\rm EEC}}^{q}(x,\theta,\phi) &= \int \frac{d\eta^-}{2\pi} e^{-ix P^+ \eta^-} \langle PS|\bar \psi_j(\eta^-) {\cal L}_n^{\dagger}(\eta^-){\cal E}(\theta,\phi) {\cal L}_n(0) \psi_i(0) |PS\rangle~. \label{eq:NEEC2}
\end{align}
This expression shows that the quark NEECs describe the correlations between the removal of a quark from the target nucleon and the formation of an energy flow from the target remnants.

To establish the connection between the correlation matrices ${\cal M}^{q}_{ij,\text{FrF}}$ and ${\cal M}_{ij,{\rm EEC}}^{q}$, we first observe that the single inclusive summation in the fracture matrix ${\cal M}^{q}_{ij,\text{FrF}}$ yields the number operator of the identified hadron~\cite{Collins:1981uw,Levelt:1993ac}: 
\begin{align}
 \sum_X\int\frac{d^3 \bs P_X}{2 E_X(2\pi)^3}|P_h X \rangle \langle X P_h| =a_{h}^\dagger a_h~,
\label{eq:sum1}
\end{align}
where $a_{h}^\dagger$ and $a_h$ denote the creation and annihilation operators of the identified hadron $h$ with momentum $P_h$, respectively. 

Similarly, the energy flow operator in the NEEC matrix ${\cal M}_{ij,{\rm EEC}}^{q}$ can be expressed in terms of the hadronic number operator as~\cite{Sveshnikov:1995vi,Bauer:2008dt}:
\begin{align}
{\cal E}(\theta,\phi)=\sum_h \int \frac{d^3 \bs P_h }{2E_h(2\pi)^3}\frac{E_h}{E_N}\delta(\theta^2-\theta^2_h)\delta(\phi-\phi_h)a^\dagger_h
a_h~,\label{eq:sum2}
\end{align}
where $\sum_h$ represents the summation over all possible hadrons in the out states. 


 By comparing the expressions in eq.~(\ref{eq:sum1}) and eq.~(\ref{eq:sum2}), we identify a sum rule that connects the correlation matrix of fracture functions to that of NEECs:
  \begin{align}
{\cal M}^{q}_{ij,\text{EEC}}(x,\theta,\phi)=&\sum_{h}\int_0^{1-x}\xi_hd\xi_h \int  d^2 \bs P_{h\perp} \delta(\theta^2-\theta^2_h)\delta(\phi-\phi_h) 
{\cal M}^{q}_{ij,\text{FrF}}(x,\xi_h,\bs P_{h\perp})~.
\label{eq:relation1}
\end{align} 
This relationship demonstrates that the NEECs can be derived from the associated fracture functions by integrating over the phase space for given angles $(\theta,\phi)$, weighted with the momentum fraction $\xi_h$ and then summing over all possible species of hadrons. Fundamentally, this sum rule is rooted in energy conservation, asserting that the total hadronic energy in the final states must equal the sum of the energies of each individual hadron species. Hence, it serves as an energy sum rule and can be interpreted as a parton-level extension of the cross-section connection provided in eq.~(\ref{eq:energy pattern}). This suggests that fracture functions can be regarded as the parent functions of NEECs.

We can apply the small-$\theta$ approximation to evaluate the angular constraints in eq.~(\ref{eq:relation1}). By integrating over the transverse momentum $\bs P_{h\perp}$, we obtain a compact formula:
\begin{align}
 {\cal M}_{ij,{\rm EEC}}^{q}(x,\theta,\phi) 
=\sum_{h} \int_0^{1-x} d\xi_h~ \xi_h\frac{\bs P_{h\perp}^2}{2\theta^2} {\cal M}^{q}_{ij,\text{FrF}}(x,\xi_h,\bs P_{h\perp})
\Big\vert_{\substack{ \bs P_{h\perp} = \frac{ \xi_h\theta P^+ }{\sqrt{2}}}\bs n_t}~,
\label{eq:relation2}
\end{align}
where $\bs n_t\equiv (\cos\phi,\sin\phi)$ is a unit vector in the azimuthal plane. Here and in eq.~(\ref{eq:relation1}), the upper limit of the integral over $\xi_h$ are imposed by the fracture functions, for positivity of the energy in its out states $X$. 


Additionally, in the moment space, the sum rule can be expressed as:
\begin{align}
 {\cal M}_{ij,{\rm EEC}}^{q}(N,\theta,\phi) 
=\sum_{h} \int_0^{1} d\xi_h~ \xi_h\frac{\bs P_{h\perp}^2}{2\theta^2} {\cal M}^{q}_{ij,\text{FrF}}(N,\xi_h,\bs P_{h\perp})
\Big\vert_{\substack{ \bs P_{h\perp} = \frac{ \xi_h\theta P^+ }{\sqrt{2}}}\bs n_t}~,
\end{align}
where the Mellin moments of the correlation matrices are defined as
\begin{align}
{\cal M}_{ij,{\rm EEC}}^{q}(N,\theta,\phi)=&\int_0^1 d x~x^{N-1} {\cal M}_{ij,{\rm EEC}}^{q}(x,\theta,\phi)~,
\notag \\ 
{\cal M}^{q}_{ij,\text{FrF}}(N,\xi_h,\bs P_{h\perp})=&\int_0^{1-\xi_h} d x~x^{N-1} {\cal M}^q_{ij,\text{FrF}}(x,\xi_h,\bs P_{h\perp})~.
\end{align} 
The moment-space expressions are useful for resummation studies (e.g.,~\cite{Liu:2022wop,Cao:2023oef,Cao:2023qat}).
\subsubsection{Sum rules for the individual functions \label{sec:param}}

We proceed by decomposing the quark correlation matrices, $ {\cal M}_{ij,{\rm EEC}}^{q}$ and ${\cal M}^{q}_{ij,\text{FrF}}$ and deriving the sum rules for individual fracture functions and NEECs. For the applications in the DIS energy pattern, as will be detailed in section~\ref{sec:SFfactorization}, we focus on the chiral-even contributions up to twist-3. Extending our analysis to include other contributions is straightforward.

The sum rule in eqs.~(\ref{eq:relation1}),~(\ref{eq:relation2}) shows that the fundamental behaviors of the fracture matrix ${\cal M}^{q}_{ij,{\rm FrF}}$ under hermiticity, parity, and rotational transformations are preserved in the NEEC matrix $ {\cal M}_{ij,{\rm EEC}}^{q}$. These properties ensure that the decompositions of $ {\cal M}_{ij,{\rm EEC}}^{q}$ and ${\cal M}^{q}_{ij,\text{FrF}}$ share a similar structure, resulting in a one-to-one correspondence between the NEECs and fracture functions. Additionally, it is noted that time-reversal invariance does not constrain either the fracture functions or NEECs, due to the out-state interactions contained in those matrices, as shown in eqs.~(\ref{eq:sum1}),~(\ref{eq:sum2}). This provides accommodation for the T-odd effects in the TFR, such as SSA.


For quark fracture functions, the generic decomposition has been given in~\cite{Chen:2023wsi}:
\begin{align}
 {\cal M}^{q}_{ij,\text{FrF}}(x,\xi_h,&\bs P_{h\perp}) = \frac{\left(\gamma_\rho\right)_{ij}}{2N_c} \Bigg[ \bar n^\rho \Bigl( u_{1}^q - \frac{P_{h\perp} \cdot \tilde S_\perp}{M} u_{1T}^{h,q} \Bigr)
\notag \\ 
&+ \frac{1}{P^+}\Bigl( P_{h\perp}^{\rho} u^{h,q}  - M \tilde S_{\perp}^{\rho} u^q_{T}  - S_L \tilde P_{h\perp}^{\rho} u_{L}^{h,q} - \frac{P_{h\perp}^{\langle\rho} P_{h\perp}^{\beta\rangle}}{M} \tilde S_{\perp\beta} u_{T}^{h,q}  \Bigr)\Bigg] \nonumber\\
&+ \frac{\left(\gamma_5\gamma_\rho\right)_{ij}}{2N_c} \Bigg[ \bar n^\rho \Bigl( S_L l_{1L}^q  - \frac{P_{h\perp} \cdot S_\perp}{M} l_{1T}^{h,q}  \Bigr)
\notag \\ 
& + \frac{1}{P^+}\Bigl( \tilde P_{h\perp}^\rho l^{h,q} + M S_{\perp}^\rho l_T^q  + S_L P_{h\perp}^\rho l_{L}^{h,q}  - \frac{P_{h\perp}^{\langle\rho} P_{h\perp}^{\beta \rangle }}{M} S_{\perp\beta} l_{T}^{h,q}  \Bigr) \Bigg] + \cdots~, \label{eq:FrF-u}
\end{align}
where $\cdots$ denote the chiral-odd terms or terms beyond twist-3, and $a^{\langle\alpha}_\perp a^{\beta\rangle}_\perp \equiv a^\alpha_\perp a^\beta_\perp+g_\perp^{\alpha\beta}\bs a_\perp^2/2$. All the fracture functions, $u$'s and $l$'s, are scalar functions of $x$, $\xi_h$ and $\bs  P_{h\perp}^2$. They are all real-valued due to the constraints from hermiticity. Those with ``1" in the subscript are of twist-2, while the remaining ones are of twist-3. The ``$L$''  or ``$T$'' in the subscript denotes the dependence on the longitudinal or transverse polarization of the nucleon. The superscript ``$h$'' indicates the explicit $\bs P_{h\perp}$-dependence in the decomposition.

   The parton interpretations of the four twist-2 quark fracture functions~$
   \{u_{1}^q,l_{1L}^q, u_{1T}^{h,q}, l_{1T}^{h,q}\}$ are clear~\cite{Anselmino:2011ss}. Among them, the Sivers-type fracture function $u_{1T}^{h,q}$ and the worm-gear type function $l_{1T}^{h,q}$ are of particular interest. They are known to induce the SSA $A_{UT}^{\sin(\phi_h-\phi_S)}$ and DSA $A_{LT}^{\cos(\phi_h-\phi_S)}$ for SIDIS in the TFR~\cite{Chen:2021vby}. In contrast, the remaining eight twist-3 quark fracture functions do not have simple parton interpretation as the twist-2 fracture functions, and they are related to more general fracture functions from quark-gluon-quark correlations~\cite{Chen:2023wsi}. Despite their complexity, these functions play a crucial role in generating various spin and azimuthal asymmetries in the TFR SIDIS.

The quark correlation matrix of the NEECs in eq.~(\ref{eq:NEEC2}) is decomposed similarly to the fracture functions. The chiral-even contributions up to twist-3 are given by:
\begin{align}
&{\cal M}_{ij,{\rm EEC}}^{q}(x,\theta,\phi)= \frac{\left(\gamma_\rho\right)_{ij}}{2N_c} \bigg[ \bar n^\rho \Bigl( \frac{1}{2\pi} f_{1}^q - n_t \cdot \tilde S_\perp ~ f_{1T}^{t,q} \Bigr) \nonumber\\
&\qquad\qquad\qquad + \frac{M}{P^+} \Bigl( n_t^{\rho} ~ f^{t,q}  - \frac{1}{2\pi} \tilde S_{\perp}^{\rho} ~ f_{T}^q  - S_L \tilde n_t^{\rho} ~ f_{L}^{t,q} - n_t^{\langle \rho} n_t^{\beta\rangle} \tilde S_{\perp\beta} ~ f_{T}^{t,q}  \Bigr)\bigg] \nonumber\\
&\qquad\qquad \qquad+ \frac{\left(\gamma_5\gamma_\rho\right)_{ij}}{2N_c} \bigg[ \bar n^\rho \Bigl( \frac{1}{2\pi} S_L ~ g_{1L}^q  - n_t \cdot S_\perp ~ g_{1T}^{t,q}  \Bigr) \nonumber\\
&\qquad\qquad\qquad + \frac{M}{P^+} \Bigl( \tilde n_t^\rho ~ g^{t,q}  +  \frac{1}{2\pi} S_{\perp}^\rho ~ g_{T}^q  + S_L n_t^\rho ~ g_{L}^{t,q}  - n_t^{\langle\rho} n_t^{\beta\rangle} S_{\perp\beta} ~ g_{T}^{t,q}  \Bigr) \bigg]~.\label{eq:NEEC-f}
\end{align}
Note that the azimuthal-angle $\phi$ dependences have been contained in the transverse vector  
\begin{align}
n_t^\mu= (0,0,\bs n_t)=(0,0,\cos \phi, \sin \phi)~
\label{eq:nt}.
\end{align}
Thus, the NEECs, denoted by $f$'s and $g$'s, are rotationally invariant functions depending on the variable $x$ and the polar angle $\theta$. Meanwhile, these functions are real-valued because of hermiticity. The naming conventions for these NEECs are akin to those for fracture functions. The superscript ``$t$'' indicates the $n_t$-dependence in the decomposition. If a quark NEEC is associated with the $n_t$-even structure, it survives when the angle $\phi$ is integrated out:
\begin{align}
 {\cal M}_{ij,{\rm EEC}}^{q}(x,\theta)=&\frac{\left(\gamma_\rho\right)_{ij}}{2N_c} \Big(f_{1}^q - 
\frac{M}{P^+} \tilde S_{\perp}^{\rho} ~ f_{T}^q \Big)
+ \frac{\left(\gamma_5\gamma_\rho\right)_{ij}}{2N_c} \Big( \bar n^\rho S_L ~ g_{1L}^q  + \frac{M}{P^+}  S_{\perp}^\rho ~ g_{T}^q   \Big)~.
\label{eq:nteven}
\end{align}
For these $n_t$-even NEECs, a front factor $1/2\pi$ is included in eq.~(\ref{eq:NEEC-f}) to maintain consistency with the convention adopted in \cite{Liu:2022wop}, particularly for $f_{1}^q$.

Substituting eq.~(\ref{eq:FrF-u}) and eq.~(\ref{eq:NEEC-f}) into eq.~(\ref{eq:relation2}), we obtain individual relationships for each of the quark NEECs and fracture functions: 
\begin{align}
& f_{1}^q(x,\theta) = 2\pi \sumint u_1^q(x,\xi_h,\bs P_{h\perp}^2)~,\quad
f_{1T}^{t,q}(x,\theta) =  \sumint \frac{|\bs P_{h\perp}|}{M} u_{1T}^{h,q}(x,\xi_h,\bs P_{h\perp}^2)~,  \notag 
\\
& g_{1L}^q(x,\theta) = 2\pi \sumint l_{1L}^q(x,\xi_h,\bs P_{h\perp}^2)~,\quad
g_{1T}^{t,q}(x,\theta) =  \sumint \frac{|\bs P_{h\perp}|}{M} l_{1T}^{h,q}(x,\xi_h,\bs P_{h\perp}^2)~, \notag  \\
& f^{t,q}(x,\theta) =  \sumint \frac{|\bs P_{h\perp}|}{M} u^{h,q}(x,\xi_h,\bs P_{h\perp}^2)~,\quad 
f_{L}^{t,q}(x,\theta) =  \sumint \frac{|\bs P_{h\perp}|}{M} u_L^{h,q}(x,\xi_h,\bs P_{h\perp}^2)~,  \notag\\
& f_{T}^q(x,\theta) = 2\pi \sumint u_T^q(x,\xi_h,\bs P_{h\perp}^2)~,\quad
f_{T}^{t,q}(x,\theta) =  \sumint \frac{\bs P_{h\perp}^2}{M^2} u_T^{h,q}(x,\xi_h,\bs P_{h\perp}^2)~, \notag \\
& g^{t,q}(x,\theta) =  \sumint \frac{|\bs P_{h\perp}|}{M} l^{h,q}(x,\xi_h,\bs P_{h\perp}^2)~,\quad
g_{L}^{t,q}(x,\theta) =  \sumint \frac{|\bs P_{h\perp}|}{M} l_L^{h,q}(x,\xi_h,\bs P_{h\perp}^2)~,  \notag\\
& g_{T}^q(x,\theta) =2\pi\sumint  l_T^q(x,\xi_h,\bs P_{h\perp}^2)
~,\quad
g_{T}^{t,q}(x,\theta) =  \sumint \frac{\bs P_{h\perp}^2}{M^2} l_T^{h,q}(x,\xi_h,\bs P_{h\perp}^2)~. \label{eq:FrFrelation}
\end{align}
Here, we have defined the notation for brevity:
\begin{align}
\sumint {\cal A}
\equiv\sum_{h} \int_0^{1-x} d\xi_h~ \xi_h\frac{\bs P_{h\perp}^2}{2\theta^2}~ {\cal A} \bigg\vert_{|\bs P_{h\perp}| = \frac{\theta \xi_h P^+}{\sqrt{2}}}~.
\label{eq:sumint}
\end{align}
It is apparent that there are twelve chiral-even quark NEECs up to twist-3. They have a clear one-to-one correspondence with the quark fracture functions, as shown in eq.~(\ref{eq:FrFrelation}). As expected, four of them are at twist-2. There are an unpolarized quark NEEC $f_{1}^q$ for an unpolarized nucleon target, and a quark helicity NEEC $g_{1L}^q$ for a longitudinally polarized nucleon. For a transversely polarized nucleon, one can introduce a quark spin-independent NEEC $f_{1T}^{t,q}$, the Sivers-type NEEC, and a quark helicity NEEC $g_{1T}^{t,q}$, the worm-gear-type NEEC. These two NEECs are $n_t$-odd. Given the insights from the associated fracture functions~$\{u_{1T}^{h,q},l_{1T}^{h,q}\}$, one expects that they can generate non-trivial azimuthal asymmetries for the energy pattern in DIS~(see section~\ref{sec:asymmetry}). Besides, they manifest unique behaviors at large $\theta$~(see section~\ref{sec:matching}). Additionally, akin to the fracture functions, the eight twist-3 quark NEECs lack a simple parton interpretation. Their roles in the energy pattern will be illustrated in section~\ref{sec:twist3}.

\subsubsection{Discussions}  
\label{sec:conDiscussion}
Let us further explore the implications of the sum rule outlined in eqs.~(\ref{eq:relation1}),~(\ref{eq:relation2}),~(\ref{eq:FrFrelation}). First, the sum rule preserves essential correlations between initial and final states in the TFR, establishing a direct one-to-one correspondence between NEECs and fracture functions, as shown in eq.~(\ref{eq:FrFrelation}). Consequently, this facilitates thorough investigations into properties of NEECs through the analysis of fracture functions.

 For example, the evolution equations of NEECs can be directly derived from those of fracture functions.  Notably, while the sum rule forms an integral relationship as shown in eqs.~(\ref{eq:relation1}),~(\ref{eq:relation2}),~(\ref{eq:FrFrelation}), it does not modify their ultraviolet behaviors. In other words, NEECs and fracture functions share the same evolution kernel. This enables the extension of the sum rules from the bare functions to their renormalized counterparts within a consistent renormalization scheme. An illustration of this point is provided in appendix \ref{sec:evolution}.




Similarly, the behavior of NEECs under various kinematic limits can be analyzed. An application of this analysis is detailed in section~\ref{sec:matching}, where we derive the large-$\theta$ behaviors of quark NEECs using the associated fracture functions at large $\bs P_{h\perp}$. Additionally, while an initial examination of NEECs in the small-$x$ region has been conducted in \cite{Liu:2023aqb},  exploring this aspect through small-$x$ fracture functions is also promising. A study of the small-$x$ fracture functions is currently under preparation.

 In fact, several extensions of this sum rule can be derived. In subsection \ref{sec:gluonNEECs}, we will establish the sum rules for the gluonic contributions. This extension is viable because the relationship between the single inclusive sum in fracture functions and the energy flow operator in NEECs, as shown in eqs.~(\ref{eq:sum1}),(\ref{eq:sum2}), remains unaffected by changes in the partonic operators. A similar argument applies to multi-parton NEECs and their associated fracture functions~(see relevant discussions in section~\ref{sec:twist3}). Furthermore, in appendix~\ref{sec:semiEEC}, a generic connection is established between  multi-point NEEC~\cite{Liu:2022wop,Liu:2024kqt} and multi-hadron fracture functions~\cite{Ceccopieri:2007th}, as illustrated in eq.~(\ref{eq:relationAP1}). This analysis can also be generalized to semi-inclusive energy correlators~\cite{Liu:2024kqt} in the TFR, an example of which is given in eq.~(\ref{eq:relationAP2}). 

 This framework of sum rules provides an efficient tool for studying the energy-weighted observables that involve NEECs, through hadron production processes that entail fracture functions. By using this connection, in section~\ref{sec:SFfactorization}, we derive the factorization formulas for all the structure functions in the DIS energy pattern from those in the TFR SIDIS.

Along the same lines, the investigations into TMD NEECs are feasible. For example, the classification of TMD quark fracture functions \cite{Anselmino:2011ss} allows the introduction of corresponding TMD NEECs, with sum rules detailed in appendix~\ref{sec:TMD_NEEC}. Given that the TMD quark fracture functions adhere to the same evolution and renormalization equations as conventional TMD PDFs \cite{Chai:2019ykk,Hauksson:2024bvv},  analogous equations for TMD NEECs can be derived using the sum rules. Furthermore, as quark TMD fracture functions can be accessed through the dihadron production in SIDIS \cite{Anselmino:2011bb,Anselmino:2011vkz,Guo:2023uis}, there is potential to measure associated TMD NEECs similarly. In the rest of our manuscript, we will focus on the collinear functions.

\subsection{The connection: gluonic sector~\label{sec:gluonNEECs}}  
We turn now to the sum rules for gluonic contributions. This extension is straightforward. We first introduce the gluonic fracture functions and NEECs, respectively. Then, we present the sum rules that connect them. 
  
The collinear gluon fracture functions for observing a hadron $h$ are defined through the following correlation matrix:
\begin{align}
{\cal M}_{G,\text{FrF}}^{\alpha \beta}(x, \xi_h, \bs P_{h\perp})= & \frac{1}{2\xi_h(2\pi)^3} \frac{1}{x P^{+}} \int \frac{d \lambda}{2 \pi} e^{-i \lambda x P^{+}}\sum_X\int\frac{d^3 \bs P_X}{(2\pi)^32 E_X} 
\notag\\ 
&\times \big\langle PS\big|\big(G^{+\alpha}(\lambda n) {\cal L}_n^{\dagger}(\lambda n)\big)^a\big|P_{h} X \big
\rangle \big\langle  P_{h} X\big|\big({\cal L}_n(0) G^{+\beta}(0)\big)^a\big| PS\big\rangle~,
\label{eq:gluonMG}
\end{align}
where $\alpha$ and $\beta$ are both transverse indices. $G^{\alpha\beta}$ is the gluon strength tensor, and ${\cal L}_n (x)$ is the light-cone gauge link in the adjoint representation. The classification of the gluonic fracture functions is similar to that of the gluonic TMD PDFs~\cite{Boussarie:2023izj}.

The leading twist expansion of the matrix ${\cal M}_{G,\text{FrF}}$ contains eight gluonic fracture functions~\cite{Chen:2024brp}:
\begin{align}
{\cal M}_{G,\text{FrF}}^{\alpha \beta}= &
- \frac{1}{2}g_\perp^{\alpha \beta} u_{1}^g + \frac{1}{2M^2}\Big( P_{h\perp}^\alpha   P_{h\perp}^\beta + \frac{1}{2}g_\perp^{\alpha \beta} \bs P_{h\perp}^2\Big) t_{1}^{h,g} 
+ S_L\bigg[i\frac{\varepsilon_\perp^{\alpha \beta}}{2} l_{1L}^g
+ \frac{\tilde  { P}_{h\perp}^{\{\alpha}  P_{h\perp}^{\beta\}}}{4 M^2} t_{1L}^{h,g}\bigg] \nonumber\\
& + \frac{g_\perp^{\alpha \beta} }{2}
\frac{P_{h\perp}\cdot \tilde S_\perp }{M} u_{1 T}^{h,g} + \frac{P_{h\perp} \cdot S_\perp}{M} \bigg[ i\frac{\varepsilon_\perp^{\alpha \beta}}{2} l_{1T}^{h,g} 
- \frac{\tilde P_{h\perp}^{\{\alpha} P_{h\perp}^{\beta\}}}{4M^2} t_{1T}^{hh,g} \bigg]
\notag \\ 
&
+\frac{\tilde P_{h\perp}^{\{\alpha} S_\perp^{\beta\}}+ \tilde S_\perp^{\{\alpha} P_{h\perp}^{\beta\}}}{8 M} t_{1T}^{h,g}~,
\label{eq:gluonFrF}
\end{align} 
where the notation $a^{\{\alpha} b^{\beta\}
 } \equiv a^\alpha b^\beta+a^\beta b^\alpha$ is used. All the gluonic fracture functions depend on the variables $x$, $\xi_h$ and $\bs P_{h\perp}^2$, and they are real-valued functions satisfying constraints from parity and hermiticity.

The gluonic NEECs for measuring an energy flow in the specific angle~$(\theta,\phi)$ are defined through the correlaton matrix:
\begin{align}
&{\cal M}_{G,\text{EEC}}^{\alpha \beta} (x,\theta,\phi)
\notag \\&~~~~~~= 
\frac{1}{x P^{+}} \int \frac{d \lambda}{2 \pi } e^{-i \lambda x P^{+}}\big\langle PS\big|(G^{+\alpha}(\lambda n) {\cal L}_n^{\dagger}(\lambda n))^a{\cal E}(\theta,\phi)  \big({\cal L}_n(0) G^{+\beta}(0)\big)^a\big| PS\big\rangle~,
\label{eq:gluonNEEC1}
\end{align}
where ${\cal E}( \theta,\phi) $ is the energy flow operator defined in eq.~(\ref{eq:Eflow}). Similar to eq.~(\ref{eq:gluonFrF}), there are generally eight gluonic NEECs in the matrix ${\cal M}_{G,\text{EEC}}^{\alpha \beta}$ at leading twist:
\begin{align}
{\cal M}_{G,\text{EEC}}^{\alpha \beta}= &
- \frac{1}{4\pi}g_\perp^{\alpha \beta} f_{1}^g +\Big( n_{t}^\alpha  n_{t}^\beta +\frac{1}{2}g_\perp^{\alpha \beta} \Big) h_{1}^{t,g} 
+ S_L\bigg[i\frac{\varepsilon_\perp^{\alpha \beta}}{4\pi} g_{1L}^g
+ \frac{\tilde  n_{t}^{\{\alpha}  n_{t}^{\beta\}}}{2 } h_{1L}^{t,g}\bigg] \nonumber\\
& + \frac{g_\perp^{\alpha \beta} }{2}
n_{t}\cdot \tilde S_\perp f_{1 T}^{t,g} +n_{t}\cdot S_\perp \bigg[ i\frac{\varepsilon_\perp^{\alpha \beta}}{2} g_{1T}^{t,g} 
- \tilde n_{t}^{\{\alpha} n_{t}^{\beta\}} h_{1T}^{tt,g} \bigg]
\notag \\ 
&
+\frac{\tilde n_{t}^{\{\alpha} S_\perp^{\beta\}}+ \tilde S_\perp^{\{\alpha} n_{t}^{\beta\}}}{4 } h_{1T}^{t,g}~,
\label{eq:gluonNEEC2}
\end{align}
where the azimuthal vector $n_t^\mu$ is given in eq.~(\ref{eq:nt}). The gluonic NEECs, denoted by $f$'s, $g$'s and $h$'s, are real-valued functions of the momentum fraction $x$ and the calorimeter polar angle $\theta$. They are rotationally invariant in the azimuthal plane.

Following the approach in section~\ref{sec:quarkNEEC}, one can demonstrate that the gluonic correlation matrices of the NEECs and fracture functions are connected by the following sum rule:
\begin{align}
{\cal M}_{G,\text{EEC}}^{\alpha \beta}(x, \theta,\phi) 
=&\sum_{h} \int_0^{1-x} d\xi_h~ \xi_h\frac{\bs P_{h\perp}^2}{2\theta^2}{\cal M}_{G,\text{FrF}}^{\alpha \beta}(x,\xi_h,\bs P_{h\perp})
\Big\vert_{\substack{ \bs P_{h\perp} = \frac{ \xi_h\theta P^+ }{\sqrt{2}}\bs n_t}}~, 
\label{eq:relation_gluon}
\end{align}
where the sum is over all types of hadron $h$. This connection leads to the sum rules for individual gluonic NEEC and fracture function:
\begin{align}
f_{1}^g(x,\theta)=& 2\pi\sumint  u_{1}^g(x,\xi_h,\bs P_{h\perp}^2)~,\quad\qquad~~~  h^{t,g}_{1}(x,\theta)=\sumint    \frac{|\bs P_{h\perp}|^2}{2M^2}t^{h,g}_{1}(x,\xi_h,\bs P_{h\perp}^2)~,
\notag \\ 
g_{1L}^g(x,\theta)=&2\pi \sumint  l_{1L}^g(x,\xi_h,\bs P_{h\perp}^2)~,
\quad\quad~~~~~
 h_{1L}^{t,g}(x,\theta)= \sumint \frac{|\bs P_{h\perp}|^2}{2M^2}  t_{1L}^{h,g} (x,\xi_h,\bs P_{h\perp}^2)~,\quad
 \notag \\ 
 f_{1 T}^{t,g} (x,\theta)=&\sumint   \frac{|\bs P_{h\perp}|}{M}u_{1 T}^{t,g}(x,\xi_h,\bs P_{h\perp}^2)~,~~~~
 g_{1T}^{t,g} (x,\theta)=\sumint   \frac{|\bs P_{h\perp}|}{M}l_{1 T}^{t,g}(x,\xi_h,\bs P_{h\perp}^2)~,
 \notag \\  
h^{t,g}_{1T}(x,\theta)=&  \sumint \frac{ |\bs P_{h\perp}|}{2M}
t^{h,g}_{1T}(x,\xi_h,\bs P_{h\perp}^2)~,
~~~~
 h^{tt,g}_{1T}(x,\theta)= \sumint 
\frac{ |\bs  P_{h\perp}|^3}{4M^3}t^{hh,g}_{1T}(x,\xi_h,\bs P_{h\perp}^2)~. 
\label{eq:FrFrelationG}
\end{align}
Here, the notation $\sumint$, which stands for the operations including summing over $h$ and integrating over $\xi_h$, has been defined in eq.~(\ref{eq:sumint}).

%
The sum rules in eq.~(\ref{eq:FrFrelationG}) connect eight pairs of the gluonic NEECs and fracture functions. Among them, four pairs are T-even.  The pairs $\{f_1^g,u_1^g
\}$ describe unpolarized gluons in an unpolarized nucleon, while $\{h_1^{t,g},t_1^{h,g}\}$ accommodate linearly polarized gluons. Additionally, $\{g_{1L}^g,l_{1L}^g\}$ characterize  circularly polarized gluons in a longitudinally polarized nucleon, whereas $\{g_{1T}^g,l_{1T}^g\}$ do so in a transversely polarized nucleon.

The remaining four pairs are T-odd. For example, $\{f_{1T}^{t,g}, u_{1T}^{t,g}\}$ are the Sivers-type gluon NEECs and fracture functions, respectively, describing unpolarized gluons in a transversely polarized nucleon. The other three pairs,  $\{ h_{1L}^{t,g},t_{1L}^{h,g} \}$, $\{h^{t,g}_{1T},t^{h,g}_{1T}\}$ and $\{h^{tt,g}_{1T},t^{hh,g}_{1T}\}$, characterize the gluons with tensor polarizations. These polarized gluons play a unique role in generating azimuthal asymmetries in SIDIS within the TFR~\cite{Chen:2024brp}. In section~\ref{sec:twist2}, we will observe similar effects in the DIS energy pattern. Particularly, the T-odd gluon NEECs in a transversely polarized nucleon, $h^{t,g}_{1T}$ and $h^{tt,g}_{1T}$, give rise to the SSAs of the energy pattern.

\section{Structure functions of the DIS energy pattern } 

\label{sec:SF}

In this section, we present the SFs of the energy pattern cross section in DIS. As introduced in section~\ref{sec:Introduction}, we consider the hadronic energy distribution measured in the polarized DIS process denoted as $ e(l,\lambda_e)+N(P,S)\rightarrow e(l')+X~.
$
 Here, $l, l'$ are the four momenta of the initial and final electron, respectively, and $\lambda_e$ is the lepton helicity. $P$ is the momentum of the nucleon $N$ with the polarization vector $S^\mu$. We work in the lowest order of quantum electrodynamics, where one virtual photon is exchanged between the electron and the nucleon, carrying a momentum $q=l-l'$. The conventional DIS variables are introduced by
\begin{align}
Q^2=-q^2~,\quad x_B=\frac{Q^2}{2P\cdot q}~,\quad y=\frac{P\cdot q}{P\cdot l}~.
\end{align} 
As shown in figure~\ref{fig:kinematics}, we take a reference frame in which the nucleon is fast moving along the positive $z$-direction, while the virtual photon is in the opposite direction. They carry the momenta 
\begin{align}
P^\mu\approx( P^+,0,0,0),\quad
q^\mu =\big(-x_B P^+,~ \frac{Q^2}{2x_BP^+}, 0,0\big)~.
\end{align}
The $x$- and $y$-axis are chosen in the way that the incoming lepton has the momentum:
\begin{align}
l^\mu=\Big(\frac{1-y}{y}x_B P^+, \frac{Q^2}{2x_ByP^+}, \frac{Q\sqrt{1-y}}{y},0\Big)~.
\end{align}
The spin vector of the target nucleon, as defined in eq.~(\ref{eq:spinvector}), is characterized by the nucleon helicity $S_L$ and the transverse polarization vector $\bs S_\perp=|\bs S_\perp|(\cos\phi_S,\sin\phi_S)$. The azimuthal angle $\phi_S$ is measured with respect to the lepton plane, spanned by the incoming and outgoing leptons. Additionally, we use $\psi$ to represent the azimuthal angle of the outgoing electron $e$ around the lepton beam axis relative to a reference direction. When dealing with a transversely polarized nucleon, we align this reference direction with the direction of $\bs {S}_\perp$. We then have $d\psi \approx d\phi_S$~\cite{Diehl:2005pc} in the large-$Q^2$ region.

According to eq.~(\ref{eq:energy pattern}), the differential form of the energy pattern cross section can be expressed as
\begin{align}
\frac{ d\Sigma(\theta,\phi) }{d x_B d y  d\psi }=&\sum_h \int d \xi_h d^2 \bs P_{h\perp}  \frac{d\sigma^{e+N \rightarrow e+h+X}}{d x_B d y   d\psi d \xi_h d^2 \bs P_{h\perp} }\frac{E_h}{E_N}\delta(\theta^2-\theta_h^2) \delta(\phi-\phi_h)~.
\label{eq:relationcross_section1}
\end{align} 
In the one-photon approximation, the SIDIS differential cross section has the following form~(see e.g., \cite{Chen:2023wsi,Anselmino:2011ss}):
\begin{align} 
 &\frac{d\sigma^{e+ N \rightarrow e+h+X}}{d x_B d y   d\psi d \xi_h d^2 \bs P_{h\perp}}
  =\frac{\alpha^2y}{Q^4}   L_{\mu\nu} \sum_X \int\frac{d^3 \bs P_X}{(2\pi)^32 E_X}\int \frac {d^4 x}{4\xi_h(2\pi)^4} e^{iq\cdot x} \langle P \vert J^\mu (x) \vert h X\rangle  \langle X h \vert J^\nu (0) \vert P\rangle~,
\end{align} 
where $\alpha=e^2/4\pi$ is the fine structure constant. Then, using the energy flow operator ${\cal E}(\theta,\phi)$ given in eq.~(\ref{eq:Eflow}),
the energy pattern cross section can be expressed as:
\begin{align} \frac{ d \Sigma(\theta,\phi)}{d x_B d y   d\psi }
=&\frac{\alpha^2y}{Q^4}  L_{\mu\nu}(l,\lambda_e,l') W^{\mu\nu}(q, P, S,\theta,\phi)~.
 \label{eq:energycross-section}
\end{align}
Here, the leptonic tensor is given by
\begin{align}
L_{\mu\nu}(l,\lambda_e,l')=&
 2 l_\mu l'_\nu+2 l_\nu l'_\mu-Q^2 g_{\mu\nu}+2 i \lambda_e \varepsilon_{\mu \nu\rho\sigma}q^\rho l^\sigma~,
\label{eq:leptontensor}
\end{align} 
where the lepton mass is neglected. The hadronic tensor is defined by
\begin{align}
W^{\mu\nu}(q, P, S,\theta,\phi)=\int \frac {d^4 x}{4\pi} e^{iq\cdot x} \langle PS \vert J^\mu (x){\cal E}(\theta,\phi)J^\nu (0) \vert PS\rangle~,
\end{align}
with the electromagnetic current $J^\mu=\sum_{q,\bar q} e_q \bar \psi_q\gamma^\mu \psi_q$.

It is useful to parametrize the energy pattern cross section in eq.~(\ref{eq:energycross-section}) with a set of independent SFs, based on the azimuthal modulations and the polarizations of the beam and target. As shown in~\cite{Meng:1991da}, this can be systematically achieved by performing a tensor decomposition on the hadronic tensor $W^{\mu\nu}(q, P, S,\theta,\phi)$  and subsequently contracting it with the lepton tensor. As a result, the DIS energy pattern cross section takes the following form:
\begin{align}
&\frac{ d \Sigma(\theta,\phi)}{d x_B d y   d\psi }=\frac{\alpha^2 }{x_By Q^2} \Bigg\{ \frac{A(y) }{2\pi}{ \Sigma}_{UU,T} +\frac{ E(y)}{2\pi}\Sigma_{UU,L} + B(y)\Sigma_{UU}^{\cos\phi} \cos\phi + E(y)\Sigma_{UU}^{\cos2\phi} \cos2\phi \nonumber\\
&~~~ + \lambda_e D(y)\Sigma_{LU}^{\sin\phi} \sin\phi + S_L \bigg[ B(y)\Sigma_{UL}^{\sin\phi} \sin\phi + E(y)\Sigma_{UL}^{\sin2\phi} \sin2\phi \bigg] 
\notag \\ 
&~~~+ \lambda_e S_L \bigg[ \frac{C(y)}{2\pi}\Sigma_{LL} + D(y)\Sigma_{LL}^{\cos\phi} \cos\phi\bigg] \nonumber\\
& ~~~+ |\bs S_\perp| \bigg[ \bigl( A(y)\Sigma_{UT,T}^{\sin(\phi-\phi_S)} + E(y)\Sigma_{UT,L}^{\sin(\phi-\phi_S)} \bigr) \sin(\phi-\phi_S) + E(y)\Sigma_{UT}^{\sin(\phi+\phi_S)} \sin(\phi+\phi_S) \nonumber\\
&~~~+\frac{ B(y)}{2\pi}\Sigma_{UT}^{\sin\phi_S} \sin\phi_S + B(y)\Sigma_{UT}^{\sin(2\phi-\phi_S)} \sin(2\phi-\phi_S) + E(y)\Sigma_{UT}^{\sin(3\phi-\phi_S)} \sin(3\phi-\phi_S)\bigg] \nonumber\\
& ~~~+ \lambda_e |\bs S_\perp| \bigg[ \frac{D(y)}{2\pi}\Sigma_{LT}^{\cos\phi_S} \cos\phi_S + C(y)\Sigma_{LT}^{\cos(\phi-\phi_S)} \cos(\phi-\phi_S) 
 \nonumber\\
 &~~~+ D(y)\Sigma_{LT}^{\cos(2\phi-\phi_S)} \cos(2\phi-\phi_S) \bigg] \Bigg\}~,
\label{eq:SF_EEC}
\end{align}
where we have defined the functions of the inelasticity $y$:
\begin{align}
&A(y) = y^2-2y+2~,\quad B(y) = 2(2-y)\sqrt{1-y}~, \quad  C(y) = y(2-y)~, 
\notag \\ 
&D(y) = 2y\sqrt{1-y}~, \quad E(y) = 2(1-y)~. 
\end{align}


The DIS energy pattern is described by eighteen functions, analogous to the eighteen structure functions of SIDIS. We call them as \emph{energy-pattern structure functions}~(EPSFs). All the EPSFs in eq.~(\ref{eq:SF_EEC}) are functions of the variables $x_B$, $Q^2$ as well as the polar angle  $\theta$ of the calorimeter. The first and second subscripts of these EPSFs denote the polarizations of the electron beam and the nucleon target, respectively. If present, the third subscript specifies the polarization of the virtual photon. In addition, the EPSFs with no $\phi$-modulations are normalized  by:\begin{align}
\int^{2\pi}_0 d\phi\frac{ d \Sigma(\theta,\phi)}{d x_B d y   d\psi }=&\frac{\alpha^2 }{x_By Q^2} \bigg\{ A(y) { \Sigma}_{UU,T} +E(y) \Sigma_{UU,L} + \lambda_e S_L C(y)\Sigma_{LL} 
\notag \\ 
&~~~~~
+ B(y)\Sigma_{UT}^{\sin\phi_S} \sin\phi_S+ \lambda_e |\bs S_\perp| D(y)\Sigma_{LT}^{\cos\phi_S} \cos\phi_S \bigg\}~. 
\end{align}

Furthermore, using the relation in eq.~(\ref{eq:relationcross_section1}), one can build up an one-to-one corresponding between the EPSFs and SIDIS SFs. Take $\Sigma_{UU}^{\cos\phi}$ for an example, we have  
 \begin{align}
\Sigma_{UU}^{\cos\phi}(x_B,Q^2,\theta)  
=&\sum_{h} \int_0^{1-x_B} d\xi_h\frac{\xi_h\bs P_{h\perp}^2}{2\theta^2} F_{UU}^{\cos\phi_h}   (x_B,Q^2, \xi_h,\bs P_{h\perp}^2)
\Big\vert_{\substack{ | \bs P_{h\perp}| = \frac{ \xi_h \theta P^+}{\sqrt{2}}}}~,
\label{eq:relation_SF}
 \end{align}
where ${ F}_{UU}^{\cos \phi_h}$ is the unpolarized SIDIS SF associated with $\cos\phi_h$ modulation, and $\phi_h$ refers to the azimuthal angle of the detected hadron~(see e.g.,~\cite{Chen:2023wsi}). Here, we have used the small-$\theta$ approximations for the TFR. 

\section{Factorization of the DIS energy pattern in the TFR \label{sec:SFfactorization}}

The SFs of the DIS energy pattern, as classified in eq.~(\ref{eq:SF_EEC}), encapsulate various insights into the internal structures of the nucleon and their correlations with the measured energy flow. To aid understanding within perturbative QCD, in this section, we investigate the factorization of the EPSFs within the kinematic regime of $Q\gg \Lambda_{\text{QCD}}$ and the TFR. Here, the TFR is characterized by $\theta P^+\ll Q$. In this region, it is expected that these EPSFs are factorized in term of the NEECs, which were introduced in section~\ref{sec:connection}.

We derive the factorization formulas by using the sum rule between NEECs and fracture functions, as well as the connection between the EPSFs and SIDIS SFs. To elucidate our method, in section~\ref{sec:method}, we will first focus on the unpolarized and azimuthal-angle-independent EPSF, $ \Sigma_{UU,T}$ . This specific case not only serves as a robust validation of our approach, given its extensive investigation in prior works~\cite{Liu:2022wop,Cao:2023oef}, but also lays the groundwork for subsequent analyses. The factorization of other EPSFs, which remain unexplored, can be derived using a similar procedure to that of $ \Sigma_{UU,T}$. 

After providing this example, we will proceed to present the twist-2 and twist-3 contributions, which are derived from our previous works~\cite{Chen:2023wsi,Chen:2024brp} on the TFR SIDIS. 

\subsection{Methodology
\label{sec:method}}  
\subsubsection{$\Sigma_{UU,T}$ as an example}
\label{sec:method1}  
 Our starting point lies in the integral relations between the energy pattern SFs and the SIDIS SFs outlined in section~\ref{sec:SF}. For $ \Sigma_{UU,T}$, the relation takes the form:
\begin{align}
 \Sigma_{UU,T}(x_B,Q^2,\theta)
=&2\pi\sum_{h} \int_0^{1-x_B} d\xi_h\frac{\xi_h\bs P_{h\perp}^2}{2\theta^2}{ F}_{UU,T}  (x_B,Q^2, \xi_h,\bs P_{h\perp}^2)
\Big\vert_{\substack{ |\bs P_{h\perp}| = \frac{ \xi_h \theta P^+}{\sqrt{2}}}}~.
\label{eq:relationUUT}
\end{align}
Here, ${F}_{UU,T}$ represents the unpolarized and azimuthal-angle-averaged SIDIS SF.

 At high $Q^2$, the $F_{UU,T}$ within the TFR receives non-vanishing contributions at twist-2, leading to the following factorization formula~\cite{Chen:2024brp}:
\begin{align}
F_{UU,T}(x_B,Q^2,\xi_h,\bs P_{h\perp}^2) = x_B \sum_{q,\bar q} e_q^2\int_{x_B}^{1-\xi_h} \frac{dz}{z} &\Bigl[{\cal H}_q\Big(\frac{x_B}{z},\frac{Q}{\mu}\Big) u_1^q(z, \xi_h,\bs P_{h\perp}^2,\mu) 
\notag \\ 
&+ {\cal H}_g\Big(\frac{x_B}{z},\frac{Q}{\mu}\Big) u_{1}^g(z, \xi_h, \bs P_{h\perp}^2,\mu)  \Bigr]~.
\label{eq:FrFUUT}
\end{align}
Here, $u_1^q,u_{1}^g$ are the twist-2 unpolarized quark and gluon fracture functions, respectively. $\mu$ denotes the renormalization scale. ${\cal H}_q$ and ${\cal H}_g$ denote the associated hard coefficient functions, which, for $F_{UU,T}$, are known to coincide with those in unpolarized inclusive DIS to all orders of $\alpha_s$~\cite{Collins:1997sr}. Their expressions up to one loop will be given later.

By applying the above factorization formula to eq.~(\ref{eq:relationUUT}), we obtain the EPSF $\Sigma_{UU,T}$ expressed in terms of fracture functions:
\begin{align}
 \Sigma_{UU,T}&(x_B,Q^2,\theta) \notag \\ 
=&x_B \sum_{q,\bar q} e_q^2\sum_{h} \int_{x_B}^{1} \frac{dz}{z} 
 \bigg[{\cal H}_q\Big(\frac{x_B}{z},\frac{Q}{\mu}\Big) \int_{0}^{1-z} d\xi_h   ~  \frac{\xi_h\bs P_{h\perp}^2 }{2\theta^2} u_1^q(z, \xi_h,\bs P^2_{h\perp},\mu) 
\notag \\ 
~&\qquad+ {\cal H}_g\Big(\frac{x_B}{z},\frac{Q}{\mu}\Big)  \int_{0}^{1-z} d\xi_h   ~  \frac{\xi_h\bs P_{h\perp}^2 }{2\theta^2}u_{1}^g(z, \xi_h,\bs  P^2_{h\perp},\mu)  \biggr] \bigg\vert_{\substack{ |\bs P_{h\perp}| = \frac{ \xi_h \theta P^+}{\sqrt{2}}}}~.
\label{eq:UUT1}
\end{align}
 In deriving this equation, we have interchanged the order of the integrations over the variables $\xi_h$ and $z$, using $\int_0^{1-x_B} d\xi_h \int_{x_B}^{1-\xi_h} dz  =\int_{x_B}^{1} dz \int_{0}^{1-z} d\xi_h~
$. After this interchange, the kinematic constraint $0<\xi_h<1-z$, obeyed by fracture functions, is naturally satisfied. Moreover, since the hard functions do not depend on $\xi_h$, the $\xi_h$-integration now acts solely on the fracture functions. 


According to the sum rules in eqs.~(\ref{eq:FrFrelation}) and (\ref{eq:FrFrelationG}), we can perform the $\xi_h$-integrals in eq.~(\ref{eq:UUT1}) and transform all the involved fracture functions into the associated NEECs. 
Then, we have:
\begin{align}
f_{1}^a(z,\theta,\mu)=2\pi\int_{0}^{1-z} d\xi_h  \frac{\xi_h\bs P_{h\perp}^2 }{2\theta^2} u_1^a(z,\xi_h,\bs P_{h\perp}^2,\mu) \bigg\vert_{|\bs P_{h\perp}| = \frac{ \xi_h\theta P^+}{\sqrt{2}}}~,
\label{eq:sumrulef1}
\end{align} 
with $a=q,g$. 
Using this sum rule, we finally obtain the factorization formula of $\Sigma_{UU,T}$ in terms of unpolarized quark and gluon NEECs, $f_{1}^a$ :
\begin{align}
 \Sigma_{UU,T}(x_B,Q^2,\theta)
=x_B \sum_{q,\bar q} e_q^2\int_{x_B}^{1} \frac{dz}{z} 
& \bigg[{\cal H}_q\Big(\frac{x_B}{z},\frac{Q}{\mu}\Big)  f_{1}^q(z,  \theta,\mu) 
+{\cal H}_g\Big(\frac{x_B}{z},\frac{Q}{\mu}\Big) f_{1}^g(z, \theta,\mu)  \biggr]~.
 \label{eq:UUT2}
\end{align} 
We find that our result agrees with that previously given in~\cite{Liu:2022wop,Cao:2023oef}.

\subsubsection{Discussions}
\label{eq:SFdiscussion}
Here are a few more comments on the above approach and results: 

First, by comparing the factorization formulas in eq.~(\ref{eq:FrFUUT}) and eq.~(\ref{eq:UUT2}), it becomes evident that the energy SF $\Sigma_{UU,T}$ shares the same hard coefficients with the SIDIS SF $F_{UU,T}$. This congruence arises because the inclusive energy weighting at small $\theta$, which bridges SIDIS to the energy pattern, solely involves hadrons generated from the target fragmentation.  Within the TFR SIDIS, the hard coefficients that describe large-angle partonic scattering are distinctly separated from the fracture functions, which encapsulate the dynamics of target fragmentation.  Consequently, while the inclusive energy weighting effectively converts fracture functions into associated NEECs through the sum rule, the hard coefficients remain unaffected throughout the derivation.

Furthermore, given that both the EPSFs and SIDIS SFs are renormalization-scale invariant, the consistency of their hard coefficients implies that the NEECs must obey the same evolution equations as the associated fracture functions. This verifies our claim in section~\ref{sec:conDiscussion}. For example, the fracture functions $u_1^a$ are known to follow the standard DGLAP evolution~\cite{Berera:1995fj,Grazzini:1997ih,Chai:2019ykk}~, and hence so do the associated NEECs $f_1^a$. More discussions on the associations between their evolution equations can be found in appendix~\ref{sec:evolution}.

This analysis extends to all other EPSFs in the TFR, including those contributions from the NEECs beyond the leading twist. Thanks to the sum rule between NEECs and fracture functions, once the factorization formula of a SIDIS SF in terms of fracture functions is established, deriving the corresponding EPSF in terms of NEECs becomes straightforward.

It is worth noting that the twist-2 collinear factorization of $F_{UU,T(L)}$\footnote{Here we use the notation $F_{UU,T(L)}$ to stand for $F_{UU,T}$ and $F_{UU,L}$, similarly for $\Sigma_{UU,T(L)}$, $\Sigma_{UT,T(L)}^{\sin(\phi - \phi_S)}$ in the followings.} for the TFR SIDIS was rigorously proved to all orders of $\alpha_s$ in a seminal work~\cite{Collins:1997sr}  by Collins in the late 1990s. Our derivation in section~\ref{sec:method1} demonstrates that the factorization of the $\Sigma_{UU,T(L)}$ can be proven along the same lines as in ref.~\cite{Collins:1997sr}. The only addition is the energy-weighting in eq.~(\ref{eq:relationUUT}), which is addressed by the sum rules we established.

In fact, the bulk of the proof for $F_{UU,T(L)}$ aligns closely with that of ordinary inclusive DIS but with an important distinction. As Collins highlighted in~\cite{Collins:1997sr}, a complete proof must address the soft-gluon cancellations specifically, due to the unique correlations between the initial- and final-state interactions presented in the TFR. It is critical to demonstrate that these correlations do not trap the associated soft-gluon exchanges, which would invalidate the standard soft-gluon approximation and hinder the decoupling of these gluons. As a crucial part of the proof in~\cite{Collins:1997sr}, Collins introduced a systematic prescription of contour deformations to maintain the validity of the soft-gluon approximation. Following this approach allows one to effectively decouple a soft-gluon factor from the correlations in the TFR. Consequently, this factor remains unaffected by the energy weighting that links $F_{UU,T(L)}$ to $\Sigma_{UU,T(L)}$. Moreover, this soft factor has the same form as in ordinary inclusive DIS, thus it cancels out in the standard inclusive sum. 

 These studies show that the DIS energy pattern in the TFR is well-formulated within a collinear factorization framework, akin to ordinary inclusive DIS. Soft gluon effects are completely canceled, thus they do not give rise to any large double logarithm.  This contrasts with the TMD studies in the CFR, where the Sudakov or TMD resummation is necessary to suppress such enhancements~\cite{Li:2021txc,Kang:2023big}. 

  Furthermore, one can find that both $\Sigma_{UU,T(L)}$ and $F_{UU,T(L)}$ have the same hard coefficients as the SFs in the ordinary inclusive DIS. As we will show in the next subsection, this observation also extends to $\Sigma_{UT,T(L)}^{\sin(\phi - \phi_S)}, \Sigma_{LL}$ and $\Sigma_{LT}^{\cos(\phi - \phi_S)} $, where the latter two correspond to helicity-dependent inclusive DIS.  

In the subsequent sections, with the existing results of the TFR SIDIS in~\cite{Chen:2023wsi,Chen:2024brp}, we will directly present the final results without detailing the derivations.

\subsection{Twist-2 contributions \label{sec:twist2}}
At twist-2, a total of ten EPSFs contribute to the energy pattern in the TFR. Among them, four EPSFs begin to receive non-vanishing contributions starting from ${\cal O}(\alpha_s^0)$. We first focus on these four EPSFs. Two of these four are generated by the unpolarized quark and gluon NEECs. One is $\Sigma_{UU,T}$, as already given in eq.~(\ref{eq:UUT2}). 

The other is the Sivers-type EPSF, yielding:
\begin{align} 
& \Sigma_{UT,T}^{\sin(\phi - \phi_S)} = x_B \sum_{q,\bar q} e_q^2\int_{x_B}^1 \frac{dz}{z} \Bigl[  {\cal H}_q\Big(\frac{x_B}{z}\Big)  f_{1T}^{t,q}(z,\theta)
+{\cal H}_g\Big(\frac{x_B}{z}\Big)  f_{1T}^{t,g}(z, \theta)  \Bigr]~.
\label{eq:UTTsivers}
\end{align} 
Here, $f_{1T}^{t,q},f_{1T}^{t,g}$ represent the  Sivers-type NEECs that describe the unpolarized quarks and gluons in a transversely polarized nucleon, respectively. These functions contain T-odd effects and thus give rise to single transverse spin asymmetry, known as the Sivers-type asymmetry, in the TFR. It is interesting to find that the $\Sigma_{UT,T}^{\sin(\phi_h - \phi_S)}$ have the same hard coefficients functions with $\Sigma_{UU,T}$, expressed by:
\begin{align}
 {\cal H}_q(z) =& \delta(\bar z) + \frac{\alpha_s}{2\pi} \Biggl\{ P_{qq}(z) \ln\frac{Q^2}{\mu^2} + C_F \biggl[ 2\left( \frac{\ln\bar z}{\bar z} \right)_+ - \frac{3}{2}\left(\frac{1}{\bar z}\right)_+ 
 - (1+z)\ln\bar z 
 \notag \\
&\qquad\qquad \quad - \frac{1+z^2}{\bar z} \ln z + 3 - \left( \frac{\pi^2}{3} + \frac{9}{2} \right) \delta(\bar z) \biggr] \Biggr\}+{\cal{O}}(\alpha_s^2)~, \nonumber\\
 {\cal H}_g(z) =& \frac{\alpha_s}{2\pi} \biggl[ P_{qg}(z) \ln \frac{Q^2\bar z}{\mu^2 z}  - T_F (1-2z)^2 \biggr]+{\cal{O}}(\alpha_s^2)~, 
\label{eq:hardfunction1}
\end{align}
where $\bar z\equiv1-z$, and the splitting functions are given by:
\begin{align}
 P_{qq}(z) = C_F \left[ \frac{1+z^2}{(1- z)_+} + \frac{3}{2}\delta(1-z) \right]~, \quad
 P_{qg}(z)=T_F\left [  z^2
 +(1-z)^2\right]~.
\end{align}

The other two EPSFs that have non-zero contributions staring from ${\cal} O(\alpha_s^0)$, are given by:
\begin{align}
& \Sigma_{LL} = x_B \sum_{q,\bar q} e_q^2\int_{x_B}^1 \frac{dz}{z}
 \Bigl[  \Delta{\cal H}_q\Big(\frac{x_B}{z}\Big)  g_{1L}^{q}(z, \theta) +\Delta {\cal H}_g\Big(\frac{x_B}{z}\Big)  g_{1L}^g(z, \theta)\Bigr]~, \nonumber\\
& \Sigma_{LT}^{\cos(\phi - \phi_S)} = x_B \sum_{q,\bar q} e_q^2 \int_{x_B}^1 \frac{dz}{z} \Bigl[\Delta{\cal H}_q\Big(\frac{x_B}{z}\Big)  g_{1T}^{t,q}(z, \theta)+\Delta{\cal H}_g\Big(\frac{x_B}{z}\Big)  g_{1T}^{t,g}(z, \theta)\Bigr]~,
\label{eq:result1}
\end{align}
where  $g_{1L}^{q}, g_{1L}^g$ are the helicity-dependent quark and gluon NEECs for an unpolarized target, respectively. Similarly, $g_{1T}^{t,q}, g_{1T}^{t,g}$ are those for an transversely polarized target, namely the worm-gear NEECs. In line with their SIDIS counterparts, the above EPSFs share the same hard coefficients functions with the polarized inclusive DIS~\cite{Chen:2024brp}, given by
\begin{align}
 \Delta {\cal H}_q(z) = &\delta(\bar z) + \frac{\alpha_s}{2\pi} \biggl\{\Delta P_{qq}(z) \ln\frac{Q^2}{\mu^2} +  C_F\biggl[ (1+z^2)\left( \frac{\ln\bar z}{\bar z} \right)_+ - \frac{3}{2}\left(\frac{1}{\bar z}\right)_+ 
 \notag \\
&\qquad\qquad \quad - \frac{1+z^2}{\bar z} \ln z + 2+z - \left( \frac{\pi^2}{3} + \frac{9}{2} \right) \delta(\bar z) \biggr] \Biggr\}+{\cal{O}}(\alpha_s^2)~, \nonumber\\
 \Delta{\cal H}_g(z) =& \frac{\alpha_s}{2\pi} \biggl[\Delta P_{qg}(z) \left( \ln\frac{Q^2\bar z}{\mu^2 z} -1 \right) + 2T_F \bar z \biggr]+{\cal{O}}(\alpha_s^2)~,
\end{align}
where $ \Delta P_{qq}(z) = P_{qq}(z),~
 \Delta P_{qg}(z)=T_F(2z-1)$ are the helicity-dependent splitting functions.

 Now we turn to the remaining six of the ten twist-2 EPSFs that begin to contribute only from one loop. Among them, two are associated with longitudinal photon: 
\begin{align}
& \Sigma_{UU,L} = \frac{\alpha_s}{2\pi}x_B  \sum_{q,\bar q} e_q^2 \int^1_{x_B} \frac{d z}{z} \bigg[
 4T_F \frac{x_B}{z}\Big(1-\frac{x_B}{z}\Big) f_{1}^g(z,\theta) +2 C_F \frac{x_B}{z} f_{1}^q(z,\theta) \bigg]+{\cal O}(\alpha_s^2)~, \nonumber\\ 
& \Sigma_{UT,L}^{\sin \left(\phi-\phi_S\right)}
= \frac{\alpha_s}{2\pi} x_B \sum_{q,\bar q} e_q^2\int^1_{x_B} \frac{d z}{z}\bigg[
 4T_F \frac{x_B}{z}\Big(1-\frac{x_B}{z}\Big) f_{1T}^{t,g}(z,\theta) + 2C_F \frac{x_B}{z} f_{1T}^{t,q}(z,\theta) \bigg]+{\cal O}(\alpha_s^2)~. \label{eq:result2}
\end{align}
These two EPSFs are yielded by the same NEECs as their counterparts, $\Sigma_{UU,T}$ and $\Sigma_{UT,T}^{\sin \left(\phi-\phi_S\right)}$ in eqs.~(\ref{eq:UUT2}) and (\ref{eq:UTTsivers}). Similarly, the hard functions are identical to those of the longitudinal SF $F_L$ in ordinary inclusive DIS~\cite{Chen:2024brp}. Here, we only present the expression up to ${\cal O}(\alpha_s)$.
 
Besides those in eq.~(\ref{eq:result2}), another four EPSFs that only becomes non-zero at one loop are summarized as follows:
\begin{align}
& \Sigma_{UU}^{\cos 2\phi} =  - \frac{\alpha_s }{2\pi} x_B \sum_{q,\bar q} e_q^2  
\int^1_{x_B} \frac{d z}{z} T_F \Big(\frac{x_B}{z}\Big)^2 h^{t,g}_{1}(z,\theta)~,
 \nonumber\\
 & \Sigma_{UL}^{\sin 2\phi} = \frac{\alpha_s}{2\pi}   x_B\sum_{q,\bar q} e_q^2
\int^1_{x_B} \frac{d z}{z} T_F \Big(\frac{x_B}{z}\Big)^2 h^{t,g}_{1L}(z,\theta)~,\nonumber
 \\ 
& \Sigma_{UT}^{\sin (3\phi-\phi_S)} =\frac{\alpha_s}{2\pi}  x_B\sum_{q,\bar q} e_q^2
\int^1_{x_B} \frac{d z}{z}T_F  \Big(\frac{x_B}{z}\Big)^2 h^{tt,g}_{1T}(z,\theta)~, \nonumber
 \\ 
& \Sigma_{UT}^{\sin (\phi+\phi_S)} = \frac{\alpha_s}{2\pi}x_B\sum_{q,\bar q} e_q^2  
\int^1_{x_B} \frac{d z}{z} T_F \Big(\frac{x_B}{z}\Big)^2\Big[h^{t,g}_{1T}(z,\theta) 
+ h^{tt,g}_{1T}(z,\theta) \Big]~,
\label{eq:twist2gluonEEC}
\end{align}
where $h^{t,g}_{1},h^{t,g}_{1L}, h^{t,g}_{1T},h^{tt,g}_{1T}$ are the associated gluonic NEECs, defined in eq.~(\ref{eq:gluonNEEC2}). 


The four EPSFs in eq.~(\ref{eq:twist2gluonEEC}), each leading to a distinct azimuthal modulation, are particularly intriguing. Although these modulations receive non-zero contributions at twist-2, they can only be induced by gluonic NEECs, as contributions from quark NEECs are absent to all orders of $\alpha_s$ due to angular momentum conservation.  While quark contributions might appear when higher-twist effects are considered, our findings in subsection~\ref{sec:twist3} suggest that these would only manifest beyond twist-3. 
 

Therefore, these four EPSFs are uniquely sensitive to the gluonic dynamics within the nucleon, allowing for the measurement of associated gluonic NEECs without quark contribution interference. Specifically, the Boer-Mulders-type EPSF $\Sigma_{UU}^{\cos 2\phi}$ is driven by the linearly polarized gluon NEEC $h^{t,g}_{1}$. This $\cos 2\phi$-modulation, accessed through a single energy flow, offers a novel method to probe the linearly polarized gluons, complementing the double-energy-flow approach recently proposed in~\cite{Li:2023gkh}. Additionally,  the other three EPSFs---$\Sigma_{UL}^{\sin 2\phi}$, $\Sigma_{UT}^{\sin (3\phi-\phi_S)}$, and $\Sigma_{UT}^{\sin (\phi+\phi_S)}$---offer the first glimpse of novel T-odd gluonic NEECs, specifically $h^{t,g}_{1L}$, $h^{t,g}_{1T}$, and $h^{tt,g}_{1T}$, which have not yet been explored in existing literature. In particular, the NEECs $h^{t,g}_{1T}$ and $h^{tt,g}_{1T}$ are instrumental in generating azimuthal correlations between the energy flow and the target's transverse spin. Together with the gluon Sivers NEEC $f_{1T}^{t,g}$, they reveal the gluonic origins of single transverse spin asymmetries in the DIS energy pattern. 
%


Furthermore, these four gluonic NEECs in eq.~(\ref{eq:twist2gluonEEC}) all correspond to the gluon tensor polarizations. Similar to gluonic TMDs~\cite{Boussarie:2023izj} or fracture functions~\cite{Chen:2024brp}, observing such gluon polarizations requires introducing a transverse reference direction, independent of the target spin. While the final hadron's transverse momentum, $\bs P_{h\perp}$, provide this reference in fracture functions, the azimuthal vector $\bs n_t$ of the measured energy flow fulfills a similar role in the NEECs, as seen in eq.~(\ref{eq:gluonNEEC2})~(see also the discussions in~\cite{Li:2023gkh}). Notably, conventional inclusive DIS lacks the capability to incorporate such a reference direction. Therefore, unlike other twist-2 contributions, the four EPSFs in eq.~(\ref{eq:twist2gluonEEC}) do not have corresponding hard coefficients in inclusive DIS. Interestingly, these four EPSFs share exactly the same hard coefficients, differing only by a sign, which results from normalization differences. It would be instructive to explore whether this pattern persists beyond one loop and to investigate any underlying mechanisms responsible for this consistency.



%




\subsection{Twist-3 contributions~\label{sec:twist3}}
We turn to present the twist-3 contributions to the EPSFs. They are derived from the existing twist-3 results of the TFR SIDIS in~\cite{Chen:2023wsi}, where only the tree-level contributions are currently available. We summarize the final results as follows:
\begin{align} 
& \Sigma_{UU}^{\cos\phi} = -2\frac{M}{Q}x_B^2 \sum_{q,\bar q} e_q^2  f^{t,q}(x_B,\theta)~, \qquad\quad 
\Sigma_{UL}^{\sin\phi} = -2\frac{M}{Q}x_B^2  \sum_{q,\bar q} e_q^2  f_{L}^{t,q}(x_B,\theta)~,
\notag \\
& \Sigma_{UT}^{\sin\phi_S} = - 2\frac{M}{ Q} x_B^2 \sum_{q,\bar q} e_q^2   f_{T}^q(x_B,\theta)~, \qquad 
\Sigma_{UT}^{\sin(2\phi-\phi_S)} = -2\frac{M}{Q} x_B^2f_{T}^{t,q}(x_B,\theta)~,\notag \\
& \Sigma_{LU}^{\sin\phi} = 2\frac{M}{Q} x_B^2 \sum_{q,\bar q} e_q^2   g^{t,q}(x_B,\theta)~, \qquad\quad~ 
\Sigma_{LL}^{\cos\phi} = -2\frac{M}{Q} x_B^2 \sum_{q,\bar q} e_q^2   g_{L}^{t,q}(x_B,\theta)~, \notag \\
& \Sigma_{LT}^{\cos\phi_S} = - \frac{M}{ Q} x_B^2 \sum_{q,\bar q} e_q^2   g_{T}^q(x_B,\theta)~, \qquad 
\Sigma_{LT}^{\cos(2\phi-\phi_S)} = -\frac{M}{Q} x_B^2  \sum_{q,\bar q} e_q^2   g_{T}^{t,q}(x_B,\theta)~. \label{eq:twist3EEC}
\end{align}
Two comments can be made from these results. 
(i) All these EPSFs exhibit a scaling behavior as $M/Q$ at high $Q$, indicative of their twist-3 nature. (ii) Each EPSF is concisely expressed with a twist-3 quark NEEC defined in subsection \ref{sec:param}. Despite their simplicity, these expressions carry non-trivial implications, even at the tree level. Unlike the twist-2 contributions discussed earlier, exploring high-twist effects typically requires considering intricate contributions arising from multi-parton correlators~(see e.g., \cite{Wei:2016far,Qiu:1990xy}). To study the EPSFs at twist-3, one would, in principle, introduce the D-type and F-type quark-gluon-quark NEECs. However, as demonstrated in SIDIS, the application of the QCD equation of motion allows all the involved quark-gluon-quark fracture function to be transformed into quark fracture functions~\cite{Chen:2023wsi}. This mechanism, also applicable to the EPSFs due to the sum rules between the NEECs and fracture functions, finally results in the concise form given in eq.~(\ref{eq:twist3EEC}). 


\subsection{Comparisons between the TFR and the CFR}

\begin{table*}[htpb!!]
\centering 
\begin{tabular}{c|cc|cc}
\hline  
\hline  
Structure 
&\multicolumn{2}{c|}{TFR} &\multicolumn{2}{c}{ CFR } \\
functions&twist& order &twist  & order \\
\hline
$\Sigma_{UU,T}$                          & 2 & $ \alpha_s^0$ & 2 & $\alpha_s^0$ \\
$\Sigma_{UU,L}$                          &2 & $\alpha_s$ & 4 & $\alpha_s^0$  \\
$\Sigma_{UT,T}^{\sin(\phi - \phi_S)} $ &2 & $\alpha_s^0 $ & 2 & $\alpha_s^0$ \\
$\Sigma_{UT,L}^{\sin(\phi - \phi_S)} $ &2 & $\alpha_s $ & 4 & $\alpha_s^0$ \\
$\Sigma_{LL}$                            & 2 & $ \alpha_s^0$ & 2 & $\alpha_s^0$ \\
$\Sigma_{LT}^{\cos(\phi - \phi_S)} $& 2 & $ \alpha_s^0$ & 2 & $\alpha_s^0$ \\
$ \Sigma_{UU}^{\cos 2\phi}         $& 2 & $ \alpha_s$ & 2 & $\alpha_s^0$ \\ 
$ \Sigma_{UL}^{\sin 2\phi}         $& 2 & $ \alpha_s$ & 2 & $\alpha_s^0$ \\
$\Sigma_{UT}^{\sin (3\phi-\phi_S)} $& 2 & $ \alpha_s$ & 2 & $\alpha_s^0$ \\
$\Sigma_{UT}^{\sin (\phi+\phi_S)}  $& 2 & $ \alpha_s$ & 2 & $\alpha_s^0$ \\
$ \Sigma_{UU}^{\cos\phi}        $& 3 & $ \alpha_s^0$   & 3&$\alpha_s^0$ \\
$\Sigma_{UL}^{\sin\phi}         $& 3 & $ \alpha_s^0$   &3 & $\alpha_s^0$\\
$\Sigma_{UT}^{\sin\phi_S}       $& 3 & $\alpha_s^0$    & 3 & $\alpha_s^0$\\  
$\Sigma_{UT}^{\sin(2\phi-\phi_S)}$& 3 &$ \alpha_s^0$   & 3& $\alpha_s^0$ \\  
$\Sigma_{LU}^{\sin\phi}         $& 3 & $ \alpha_s^0$   &3& $\alpha_s^0$\\
$\Sigma_{LL}^{\cos\phi}         $& 3 & $ \alpha_s^0$   & 3 & $\alpha_s^0$ \\
$\Sigma_{LT}^{\cos\phi_S}        $& 3 & $ \alpha_s^0$  & 3& $\alpha_s^0$\\ 
$\Sigma_{LT}^{\cos(2\phi-\phi_S)}$& 3 & $ \alpha_s^0$  & 3 & $\alpha_s^0$ \\ 
\hline
\end{tabular}
\caption{Tables of the SFs of the DIS energy pattern. The ``twist" column indicates the leading contributions based on the power of $1/Q$, where twist-2 corresponds to $(1/Q)^0$ and twist-3 corresponds to $(1/Q)^1$. In the ``order" column, we denote the $\alpha_s$ order from which the structure functions begin to contribute. The results on the EPSFs in the TFR, where $\theta\ll 1$ are summarized from subsections~\ref{sec:twist2} and~\ref{sec:twist3}. The twist-2 results in the CFR, where $\pi-\theta\ll 1$, are summarized from ref.~\cite{Kang:2023big}. In addition, the hight-twist results in CFR are inferred from a corresponding SIDIS study in ref.~\cite{Bacchetta:2006tn,Wei:2016far}.  }
\label{table} 
\end{table*}

Through investigations up to twist-3, as shown in subsections~\ref{sec:twist2} and \ref{sec:twist3}, we now have described all eighteen EPSFs in the TFR with the relevant NEECs. Table~\ref{table} summarizes the basic characteristics of our findings. For comparative purposes, we also include the CFR results, derived in refs.~\cite{Li:2021txc,Kang:2023big} within the TMD framework at twist-2. While the higher-twist contributions in the CFR are not yet available, we infer some basic  properties from the associated SIDIS study~\cite{Bacchetta:2006tn,Wei:2016far} using the connection given in eq.~(\ref{eq:relationcross_section1}). In this discussion, we limit our focus in the CFR to the region where $\pi-\theta \ll 1$, which measures conventional TMD PDFs.

Let us compare the characteristics of the EPSFs in the TFR and the CFR:

(i) Each EPSF exhibits the same leading power behavior of $1/Q$ in both the TFR and CFR, except for the longitudinal photon EPSFs, $\Sigma_{UU,L}$ and $\Sigma_{UT,L}^{\sin(\phi - \phi_S)}$. In the TFR, these longitudinal EPSFs appear at twist-2 and are generated by $\alpha_s$ corrections, following a mechanism similar to that observed in the longitudinal cross section of ordinary inclusive DIS~(see eq.~(\ref{eq:result2})). In contrast, current studies suggest that these longitudinal EPSFs manifest at twist-4, starting from $\alpha_s^0$.  Investigating the potential of non-zero longitudinal photon EPSFs at twist-2 beyond order $\alpha_s^0$ in the CFR would be interesting. Moreover, future measurements comparing the ratio of the longitudinal to transverse cross sections, $\Sigma_{UU,L}/\Sigma_{UU,T}$, between the TFR and CFR could provide further valuable insights.

(ii) In the CFR, all EPSFs begin contributing from the order of $\alpha_s^0$. Similar characteristics are observed in the TFR up to twist-3, with two notable exceptions. The first exception, as previously discussed, involves the longitudinal photon SFs. The second exception pertains to the four azimuthal-associated SFs: $\Sigma_{UU}^{\cos 2\phi}$, $\Sigma_{UL}^{\sin 2\phi}$, $\Sigma_{UT}^{\sin (3\phi-\phi_S)}$, and $\Sigma_{UT}^{\sin (\phi+\phi_S)}$. These four SFs are uniquely induced by the gluonic NEECs at the order of $\alpha_s$, as detailed in eq.~(\ref{eq:twist2gluonEEC}). By contrast, within the CFR, these four EPSFs result from the so called Collins-type EEC jet function~\cite{Kang:2023big}, which is closely related to the Collins fragmentation functions and is inherently T-odd. Consequently, the T-even SF $\Sigma_{UU}^{\cos 2\phi}$ in the CFR emerges from a combination of T-odd functions, specifically the Collins EEC jet function paired with the Boer-Mulders quark TMD. Conversely, in the TFR, $\Sigma_{UU}^{\cos 2\phi}$ is characterized by a single T-even function, the linearly polarized gluon NEEC, $h^{t,g}_{1}$. Similar comparisons can be made for the other three azimuthal-associated EPSFs.

(iii) As we discussed in subsection~\ref{eq:SFdiscussion}, the EPSFs in the TFR are described by the collinear factorization, which is free from soft-gluon effects, and the NEECs follow the DGLAP evolutions. In contrast, soft-gluon radiations play a crucial role in the TMD factorization of the EPSFs in the CFR, driving the Collins-Soper evolution of TMD PDFs.


\subsection{Spin and azimuthal asymmetries~\label{sec:asymmetry}}
Given the structure functions in eq.~(\ref{eq:SF_EEC}), various azimuthal and spin asymmetries can be constructed for the DIS energy pattern. They are systematically defined as:
\begin{align}
\langle\mathcal{F}\rangle_{\mathcal{P}_e \mathcal{P}_N} \equiv &\frac{\int d \psi d \phi {\cal F} \frac{ d \Sigma(\theta,\phi)}{d x_B d y d\psi }}{\int d \psi d \phi \frac{ d \Sigma(\theta,\phi)}{d x_B d y d\psi }}~.
\end{align}
Here, $\mathcal{F}$ refers to a specific azimuthal angle modulation. The subscript $\mathcal{P}_e=U$ or $L$ represents the electron beam polarization, and $\mathcal{P}_N=U, L$ or $T$ denotes the target polarization. Recall that in our setup we have $d\psi\approx d \phi_S$ when the nucleon target is transversely polarized.
 
In this subsection, we present the contributions to the asymmetries of the DIS energy pattern in the TFR. For brevity, only the leading contributions for each asymmetry will be presented here. One can straightforwardly recover the full contributions using the results obtained in the previous subsections. Additionally, we imply the summation over the quark flavors $\sum_{q,\bar q} e_q^2$ in the following expressions.

Two azimuthal asymmetries are observed at order of $(1/Q)^0$ starting from tree level. They both depends on the nucleon transverse polarization. One is the SSA generated by the Sivers-type quark NEEC $f_{1 T}^{t,q}$:
\begin{align}
\left\langle\sin \left(\phi-\phi_S\right)\right\rangle_{U T} & = \frac{f_{1 T}^{t,q}\left(x_B,\theta\right)}{ 2f_1^q\left(x_B,\theta\right)}~.
\end{align} 
 The other one is the DSA:
\begin{align}
\left\langle\cos \left(\phi-\phi_S\right)\right\rangle_{L T} & =\frac{ C(y)}{2A(y)} \frac{~  g_{1 T}^{h,q}\left(x_B,\theta\right)}{f_1^q\left(x_B,\theta\right)}~,
\end{align}
induced by the worm-gear quark NEEC $g_{1 T}^{t,q}$.  

Four azimuthal asymmetries receive contributions at order $(1/Q)^0$ starting from one loop:
 \begin{align}
&\left\langle\cos 2\phi\right\rangle_{UU}
=-\frac{\alpha_s }{2\pi}\frac{E(y)}{2A(y)}\frac{   
\int^1_{x_B} \frac{d z}{z} T_F \Big(\frac{x_B}{z}\Big)^2 h^{t,g}_{1}(z,\theta)}{f_1^q\left(x_B,\theta\right)}~, 
\\ 
&\left\langle\sin 2\phi\right\rangle_{UL}
=\frac{\alpha_s}{2\pi} \frac{E(y)}{2A(y)}   \frac{
\int^1_{x_B} \frac{d z}{z}T_F \Big(\frac{x_B}{z}\Big)^2 h^{t,g}_{1L}(z,\theta)}{f_1^q\left(x_B,\theta\right)}~,
\\ 
&\left\langle\sin (3\phi-\phi_S)\right\rangle_{UT}
= \frac{\alpha_s}{2\pi}\frac{E(y)}{2A(y)}\frac{
\int^1_{x_B} \frac{d z}{z} T_F \Big(\frac{x_B}{z}\Big)^2 h^{tt,g}_{1T}(z,\theta)}{f_1^q\left(x_B,\theta\right)}~,
\\ 
&\left\langle\sin (\phi+\phi_S)\right\rangle_{UT}
=\frac{\alpha_s}{2\pi}\frac{E(y)}{2A(y)} \frac{
\int^1_{x_B} \frac{d z}{z}T_F \Big(\frac{x_B}{z}\Big)^2\Big[h^{t,g}_{1T}(z,\theta) 
+ h^{tt,g}_{1T}(z,\theta) \Big]}{f_1^q\left(x_B,\theta\right)}~.
\end{align}
As analyzed in section ~\ref{sec:twist2}, all these asymmetries are only generated by the gluonic NEECs up twist-3. It is also noted that they share the same photon flux factor $E(y)/A(y)$.

At order $1/Q$, eight azimuthal asymmetries are obtained, induced by the twist-3 quark NEECs at tree level~(${\cal O}(\alpha_s^0)$). Four are associated with the unpolarized or longitudinally polarized nucleon target:
\begin{align}
 & \left\langle\cos \phi\right\rangle_{U U}=-\frac{M}{Q} \frac{B(y)}{A(y)}
  \frac{x_B f^{t,q}\left(x_B,\theta\right)}{f_1^q\left(x_B,\theta\right)}~, \\ 
 & \left\langle\sin \phi\right\rangle_{L U}=\frac{M}{Q}\frac{D(y)}{A(y)} 
 \frac{x_B g^{t,q}\left(x_B,\theta\right)}{f_1^q\left(x_B,\theta\right)}~, \\ 
 & \left\langle\sin \phi\right\rangle_{U L}=-\frac{M}{Q} \frac{B(y)}{A(y)} 
 \frac{x_B f_L^{t,q}\left(x_B,\theta\right)}{f_1^q\left(x_B,\theta\right)}~, \\ 
 & \left\langle\cos \phi\right\rangle_{L L}=-\frac{M}{Q}\frac{D(y)}{A(y)} 
 \frac{x_B g_L^{t,q}\left(x_B,\theta\right)}{f_1^q\left(x_B,\theta\right)}~,
\end{align}
while the other four are related to the transversely polarized nucleon:
\begin{align}
 & \left\langle\sin \phi_S\right\rangle_{U T}=-\frac{M}{Q}\frac{B(y)}{A(y)}
  \frac{x_B f_{T}^{q}(x_B,\theta)}{f_1^q\left(x_B,\theta\right)}~, \\
  & \left\langle\cos \phi_S\right\rangle_{L T}=-\frac{M}{Q} \frac{D(y)}{A(y)}
  \frac{x_B g_T^q\left(x_B,\theta\right)}{f_1^q\left(x_B,\theta\right)}~, \\
   & \left\langle\sin \left(2 \phi-\phi_S\right)\right\rangle_{U T}=-\frac{M}{2Q}\frac{B(y)}{A(y)} \frac{x_B f_T^{t,q}\left(x_B,\theta\right)}{f_1^q\left(x_B,\theta\right)}~, \\
    & \left\langle\cos \left(2 \phi-\phi_S\right)\right\rangle_{L T}=-\frac{M}{2Q} \frac{D(y)}{A(y)} \frac{x_B g_T^{t,q}\left(x_B,\theta\right)}{f_1^q\left(x_B,\theta\right)}~.
\end{align}

\section{Matching of the quark NEECs at large $\theta$\label{sec:matching}}
 In the TFR where $\theta P^+\ll Q$, the DIS EPSFs can be factorized in terms of collinear NEECs. If the calorimetric measurement is restricted to the region $\theta P^+\sim\Lambda_{QCD}$, the NEECs would appear as entirely non-perturbative objects to be extracted in experiments. However, as the polar angle $\theta$ increases to the region $\theta P^+ \gg \Lambda_{QCD}$, the $\theta$-dependence can be calculated within perturbative QCD. Meanwhile, the NEECs can be further matched onto the conventional collinear parton correlation functions of nucleon. This matching was first investigated in ref.~\cite{Liu:2022wop,Cao:2023oef} for the unpolarized quark and gluonic NEEC $f_{1}^{q,g}$. Since these two NEECs are $ n_t$-even, the leading contributions in the large-$\theta$ region are expressed in terms of the twist-2 collinear parton distributions. Similarly, the twist-2 matching of linearly polarized gluon NEEC $h_{1}^{t,g}$ is provided in ref.~\cite{Li:2023gkh}.

 In this section, we extend the investigation to the remaining three leading-twist and chirality-even quark NEECs. They are the helicity NEEC $g_{1L}^q$, the Sivers NEEC $f_{1T}^{t,q}$ and the worm-gear NEEC $g_{1T}^{t,q}$~(see their definitions in eq.~(\ref{eq:NEEC-f})). Similar to $f_{1}^{q}$, the matching of the $ n_t$-even NEEC $g_{1L}^q$ at large $\theta$ is straightforward and can be expressed with the twist-2 helicity PDFs. 

 However, for the $n_t$-odd NEECs, $f_{1T}^{t,q}$ and $g_{1T}^{t,q}$, the matching calculations are much more complicated. First, to obtain non-vanishing contributions, one needs to perform the large-$\theta$ expansion to the subleading power. At this power, the final results will involve various twist-3 parton distributions, including the quark-gluon correlation functions~\cite{Efremov:1981sh,Efremov:1983eb,Qiu:1991pp,Qiu:1998ia,Eguchi:2006qz,Eguchi:2006mc} $T_F, T_\Delta$, as well as the three-gluon correlation functions~\cite{Ji:1992eu} $N, O$. In particular, deriving these twist-3 contributions is non-trivial, requiring the combined use of both the Ward identity and QCD equation of motion to ensure gauge invariance. Furthermore, an additional complication arises in the case of the Sivers NEEC, $f_{1T}^{t,q}$, which is responsible for T-odd effects and requires a nontrivial phase to be generated in the perturbative region.  

In this section, we circumvent these technical complexities by utilizing the connection to fracture functions proposed in section~\ref{sec:connection}. Although such a twist-3 matching has not been studied for NEECs, it has been investigated in detail for associated fracture functions~\cite{Chen:2021vby}. There, the matching formulas of all the four twist-2 and chirality-even quark fracture functions have already been derived in the region $ |\bs P_{h\perp}| \gg \Lambda_{QCD}$ at order $\alpha_s$. Then, according to the sum rules in eq.~(\ref{eq:FrFrelation}), we can derive the twist-3 matching formulas of the $n_t$-odd NEECs $f_{1T}^{t,q}$, $g_{1T}^{t,q}$ from those of the fracture functions $u_{1T}^{h,q}$, $l_{1T}^{h,q}$ in ~\cite{Chen:2021vby}, respectively. To illustrate our method, we first revisit the twist-2 matching of the NEEC $f_{1}^{q}$. Meanwhile, we provide the matching formula of $g_{1L}^{q}$. Afterward, we present the contributions to $f_{1T}^{t,q}$ and $g_{1T}^{t,q}$, respectively.

\subsection{Methodology: $f_{1}^q$ as an example}
Let us first recall the matching of the $\bs P_{h\perp}$-even unpolarized fracture function $u_{1}^q\left(x, \xi_h, \bs P^2_{h\perp}\right)$ studied in~\cite{Chen:2021vby}. For the large hadron transverse momentum $|\bs P_{h\perp}|\gg\Lambda_{QCD}$, at order $\alpha_s$, this fracture function can be factorized as follows:
\begin{align}
u_{1}^q\left(x, \xi_h, \bs P^2_{h\perp}\right)= & \int_{\frac{\xi_h}{1-x}}^1 \frac{d z}{z^2} \int^1_x d y\delta(x+\xi_h/z-y)\frac{ \alpha_s z^2 }{ 2\pi^2 \xi_h  \bs P_{h\perp}^2}
\notag   \\
&\times \bigg[ C_F \frac{x^2+y^2}{y^2} d_{h/g}(z) q(y)+T_R  \Big(1-\frac{x}{y}\Big)\Big[\frac{x^2}{y^2}+\big(1-\frac{x}{y}\big)^2\Big]d_{h/\bar{q}}(z) g(y)\bigg]~.
\label{eq:u1match}
\end{align}
Here, $q(y)$ and $g(y)$ represent the twist-2 unpolarized PDFs of the nucleon. $d_{h/g}(z)$ and $d_{h/\bar q}(z)$ denote the twist-2 parton fragmentation functions of an unpolarized hadron $h$. The lower limit of the integral over $z$ comes from the kinematic constraint $y=x+\xi_h/z<1$.

According to the relation given in eq.~(\ref{eq:FrFrelation}), the unpolarized quark NEEC $f_{1}^q(x,\theta)$ at $\theta P^+ \gg \Lambda_{QCD}$ can be computed from $u_{1}^q\left(x, \xi_h, \bs P^2_{h\perp}\right)$ at $|\bs P_{h\perp}|\gg\Lambda_{QCD}$. Then applying the matching formula in eq.~(\ref{eq:u1match}), we have:
\begin{align}
f_{1}^q(x,\theta)
=&\frac{1}{\theta^2}
\sum_{h} \int_0^{1-x} d\xi_h~   \int_{\frac{\xi_h}{1-x}}^1 d z \int^1_x d y\delta(x+\xi_h/z-y)
\notag   \\
&\times  \frac{ \alpha_s }{ 2\pi }\bigg[ C_F  \frac{x^2+y^2}{y^2}d_{h/g}(z) q(y) +T_R \Big(1-\frac{x}{y}\Big)\Big[\frac{x^2}{y^2}+\big(1-\frac{x}{y}\big)^2d_{h/\bar{q}}(z) g(y)\Big]\bigg]~.
\label{eq:f1qmatch1}
\end{align}  

It is expected that the inclusive energy summation in the energy flow operator guarantees the absence of final-state soft and collinear singularities for the NEECs in the perturbative region. Consequently, this obviates the necessity for fragmentation functions. To illustrate this point, let us first interchange the integration order between $\xi_h$ and $z$ by $\int ^{1-x}_0 d\xi_h \int_{\frac{\xi_h}{1-x}}^1 d z =\int ^{1}_0 dz \int^{(1-x)z}_0 d\xi_h$. Subsequently, by integrating with respect to $\xi_h$, we obtain: 
\begin{align}
f_{1}^q(x,\theta)=\frac{1}{\theta^2}
  \int^1_x d y    \frac{ \alpha_s }{ 2\pi }\sum_{h}\int^1_0 d zz& \bigg[ C_F  \frac{x^2+y^2}{y^2} d_{h/g}(z) q(y)
  \notag \\ 
  &+T_R \Big(1-\frac{x}{y}\Big)\Big[\frac{x^2}{y^2}+\big(1-\frac{x}{y}\big)^2\Big]d_{h/\bar{q}}(z) g(y)\bigg]~.
  \label{eq:f1qmatch2}
\end{align}  
It is known that the fragmentation functions adhere to the momentum sum rule~\cite{Collins:1981uw}:
\begin{align}
\sum_{h}\int_0^1 d z z d_{h/a}(z)=1~.
\end{align}
By applying this formula, we naturally eliminate all the involved fragmentation functions in the matching formula of eq.(\ref{eq:f1qmatch2}). Consequently, we arrive at the following expression:
\begin{align}
f_{1}^q(x,\theta)=&\frac{1}{\theta^2}
  \int^1_x \frac{d y}{y}  \frac{ \alpha_s }{ 2\pi }\bigg[ C_F  \Big(1+\frac{x^2}{y^2}\Big)yq(y)+T_R \Big(1-\frac{x}{y}\Big)\Big[\frac{x^2}{y^2}+\big(1-\frac{x}{y}\big)^2\Big] yg(y)\bigg]~.
\end{align}  
Our result aligns with the moment-space expression provided in~\cite{Liu:2022wop}.

Similarly, we can derive the matching formula of the helicity NEEC $g_{1L}^q(x,\theta)$ from that of the fracture function $l_{1L}^q(x,\xi_h,\bs P_{h\perp}^2)$ given in~\cite{Chen:2021vby}. The final result is expressed as:
\begin{align}
g_{1L}^q(x,\theta) = \frac{1}{\theta^2} \int_x^1 \frac{dy}{y} \frac{\alpha_s}{2\pi}\bigg[  C_F \Big(1+\frac{x^2}{y^2}\Big)y \Delta q(y) + T_R \Big(1-\frac{x}{y}\Big)\Big(\frac{2x}{y}-1\Big) y \Delta g(y) \bigg]~,
\end{align}
where $\Delta q(y)$ and $\Delta g(y)$ represent the twist-2 quark and gluon helicity PDFs, respectively.

\begin{figure}[htpp]
    \centering

    \subfigure[Hard pole]{
  \includegraphics[scale=0.45]{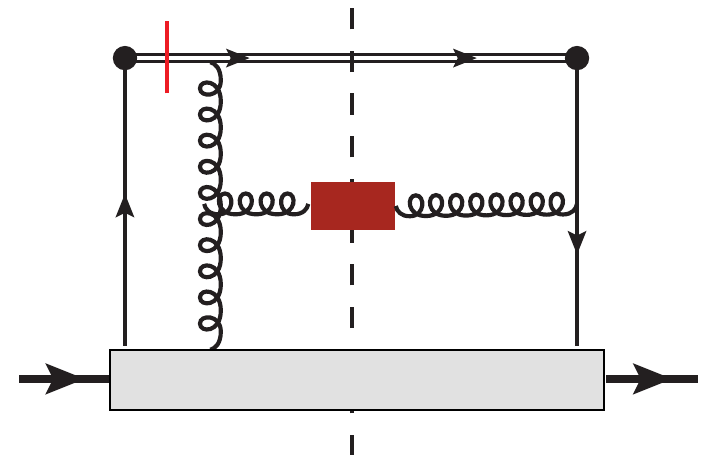}
}
\subfigure[Soft fermion pole]{
  \includegraphics[scale=0.45]{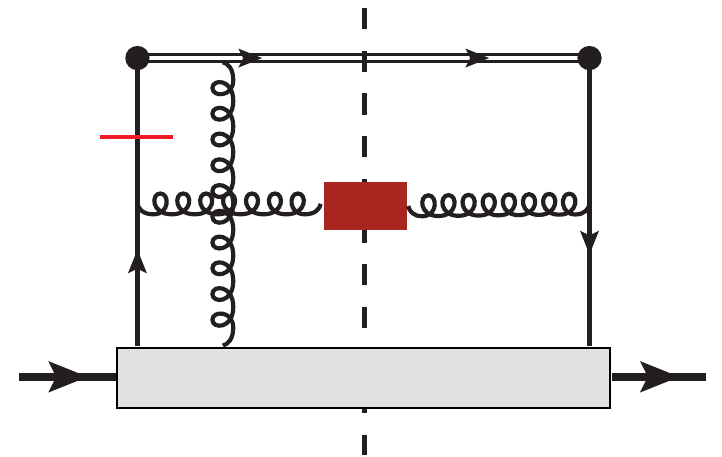}
}

\subfigure[Soft-gluon pole~($qG\bar q$)]{
  \includegraphics[scale=0.45]{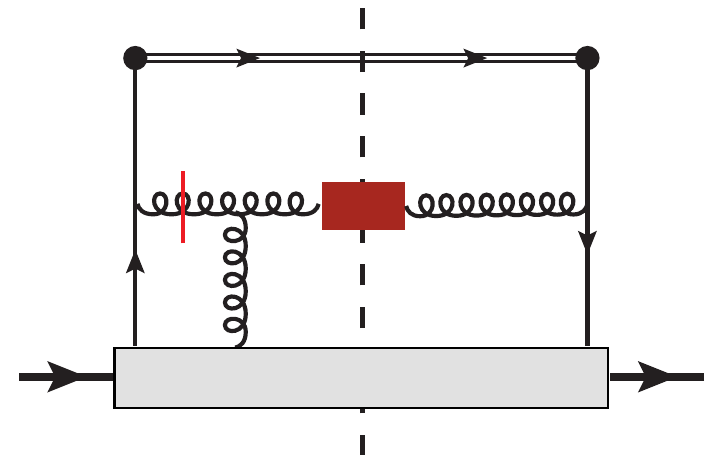}
}
\subfigure[Soft-gluon pole~($3G$)]{
  \includegraphics[scale=0.45]{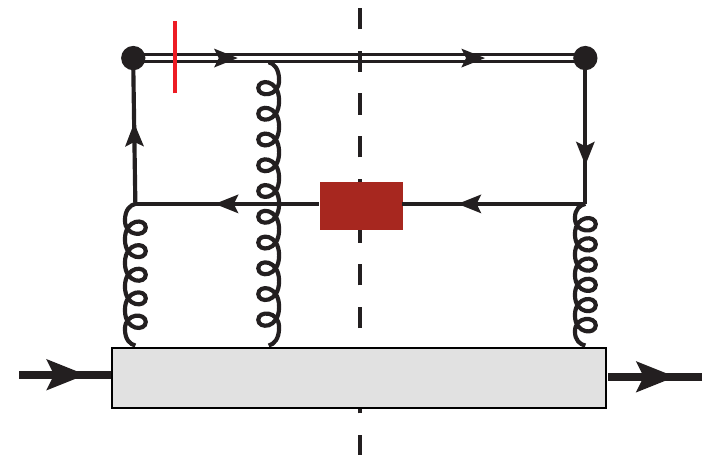}
}
  
    \caption{Typical diagrams of the Sivers-type NEEC. A short bar is used to denote the pole of the propagator under consideration. The complete set of diagrams can be found in \cite{Chen:2021vby}.}
    \label{fig:pole}
\end{figure}

\subsection{Matching of the Sivers-type NEEC}  
In this subsection, we present the complete contributions of the Sivers-type NEEC, $f_{1T}^{t,q}$, at large $\theta$. These results are derived from a study of the associated fracture function $u_{1T}^{h,q}$ in~\cite{Chen:2021vby} with the relationship outlined in eq.~(\ref{sec:asymmetry}) and the methodology introduced in the last subsection. Analogous to the fracture function $u_{1T}^{h,q}$, this NEEC characterizes the correlation between the transverse spin of the target and the orbital motion of the target fragments constituting the energy flow. Much like in SIDIS, this correlation gives rise to a single transverse spin asymmetry $\left\langle\sin \left(\phi-\phi_S\right)\right\rangle_{U T}$ within the energy pattern in the TFR, as shown in section \ref{sec:asymmetry}.  This asymmetry is particularly intriguing due to its status as a T-odd effect. We note that an illustration of the T-odd effects for the Sivers-type NEEC in the non-perturbative region has recently been provided in \cite{Liu:2024kqt}.  

In the perturbative region $\theta P^+\gg \Lambda_{\text{QCD}}$, the presence of T-odd effects entails non-zero phases (absorptive parts) in the scattering amplitude. At the level of perturbative diagrams, these phases are provided by the poles of the propagators. Given that the relationship between fracture functions and NEECs outlined in eq. (\ref{sec:asymmetry}) works diagram by diagram as well, we can obtain the various pole-contributions to the NEEC individually from~\cite{Chen:2021vby}. These contributions include the hard pole, the soft-fermion pole, and the soft-gluon pole. Representative diagrams illustrating these contributions are presented in figure \ref{fig:pole}, where the cuts on the propagators denote the positions of the poles. 
Here, we use the $F$-type correlators in the calculations, and the definitions of the these correlators are given in appendix \ref{sec:twist3PDF}.

Consequently, the matching for $f_{1T}^{t,q}(x,\theta)$ can be summarized as:\begin{align}
f_{1T}^{t,q}(x,\theta) = f_{1T}^{t,q}(x,\theta) \Big\vert_{\text{HP}} + f_{1T}^{t,q}(x,\theta) \Big\vert_{\text{SFP}} + f_{1T}^{t,q}(x,\theta) \Big\vert_{\text{SGP},qG\bar q} + f_{1T}^{t,q}(x,\theta) \Big\vert_{\text{SGP,}3G}.
\end{align}
The expressions for these four parts are given by
\begin{align}
 &f_{1T}^{t,q}(x,\theta) \Big\vert_{\text{HP}}
 = \frac{ \alpha_s N_c}{2(2\pi)^2 \theta^3 E_N} \int_x^1 \frac{dy}{y} \Bigl[ T_{\Delta }(y,x) - \bigl(1 + 2 x/(y-x) \bigr) T_F (y,x) \Bigr]~,\label{eq:f1Tt-HP} \\
  &f_{1T}^{t,q}(x,\theta) \Big\vert_{\text{SFP}} =\frac{ \alpha_s }{2(2\pi)^2 N_c \theta^3 E_N} \int_x^1 dy \frac{x}{y^3} \bigl[ (2x-y) T_F(y,0) - y T_\Delta (y,0) \bigr]~, \label{eq:f1Tt-SFP} \\
 &f_{1T}^{t,q}(x,\theta) \Big\vert_{\text{SGP},qG\bar q}  =\frac{\alpha_s N_c}{2(2\pi)^2 \theta^3 E_N} \int_x^1 \frac{dy}{y^3} \biggr [ \frac{1}{y-x} (y^3 + 3x^2 y - 2x^3)  T_F (y,y) 
\notag \\ 
&\hspace*{6cm}- y(y^2 + x^2) \frac{ d T_F (y,y) }{d y}  \biggr]~,\label{eq:f1Tt-SGP} \\
  &f_{1T}^{t,q}(x,\theta) \Big\vert_{\text{SGP,}3G} = \frac{\alpha_s }{2\pi \theta^3 E_N} \int_x^1 dy \frac{y-x}{y^5} \bigg \{ 2 (4 x^2-3 x y+y^2) \bigr[ N(y,y) - O(y,y) \bigr]
 \notag \\ 
 & \hspace*{2.5cm}- 2 (8 x^2-5 x y+y^2) \bigr[ N(y,0 ) + O(y,0) \bigr]
 + y (2x-y)^2 \frac{d}{dy} \bigr[ N (y,0)+ O(y,0) \bigr]
  \nonumber\\&
  \hspace*{2.5cm}- y (2x^2 + y^2 - 2xy)  \frac{d}{dy} \bigr[ N (y,y) - O(y,y) \bigr] \bigg \}~.\textbf{} \label{eq:f1Tt-3G}
\end{align}

\subsection{Matching of the worm-gear-type NEEC }
In contrast to the Sivers-type NEEC, the worm-gear-type NEEC $g_{1T}^{t,q}$ accounts for T-even effects. As shown in section~\ref{sec:asymmetry}, it generates a DSA $\langle\cos \left(\phi-\phi_S\right)\rangle_{LT}$ for the energy pattern in the TFR. According to eq.~(\ref{eq:FrFrelation}), this NEEC $g_{1T}^{t,q}$ corresponds to the fracture function $l_{1T}^{h,q}$ studied in \cite{Chen:2021vby}, which generates a similar asymmetry for SIDIS. It shows that unlike the single transverse spin asymmetry, this asymmetry is not zero in the absence of absorptive parts in the scattering amplitude. The typical contributions are illustrated by the diagrams in figure \ref{fig:pole} without the cut on the propagators. Moreover, it has contributions from the two-parton correlators.

From the result given there, we can summarize the matching formula of $g_{1T}^{t,q}(x,\theta)$ as follows:
\begin{align}
g_{1T}^{t,q}(x,\theta) = g_{1T}^{t,q}(x,\theta) \Big\vert_{q\bar q + qG\bar q} + g_{1T}^{t,q}(x,\theta)\Big\vert_{2G+3G}.
\end{align}
The first part is from the contribution of the quark-quark and quark-gluon-quark correlations.
From eq.~(4.12) of~\cite{Chen:2021vby}, it is given by
\begin{align}
g_{1T}^{t,q}&(x,\theta)  \Big\vert_{q\bar q + qG\bar q} 
\notag \\ 
= &\frac{\alpha_s}{(2\pi)^2 \theta^3 E_N} \int_x^1 \frac{dy}{y} \Bigg\{ 2C_F \frac{x^2}{y} q_T (y) 
  - 2C_F  \frac{ y^2 + 2 x^2 }{y^2} q_\partial (y) 
\nonumber\\
& + \frac{1}{\pi}  \int \frac{dx_2} {x_2 (y-x_2)} \biggl [ C_A\frac{x^2-x_2 y}{x_2-x} + 2 C_F \bigl(xy -x_2(x+y)\bigr)/y \biggr ] T_F (y, x_2) \nonumber\\
& - \frac{1}{\pi}  \int \frac{dx_2} {x_2 (y-x_2)} 
  \biggl [ C_A \frac{(x^2 +x_2 y)(x_2+y-2x) }{(x_2-x)(x_2-y)}  + \frac{2 C_F}{y}  \bigl(x (2x-y) + x_2 (x+y) \bigr) \biggr] T_\Delta (y,x_2) \Bigg \}~,
\end{align}
The second part is from the contribution of two-gluon and three-gluon correlations.
From eq.~(4.26) of~\cite{Chen:2021vby}, it is given by
\begin{align}
g_{1T}^{t,q}&(x,\theta)  \Big\vert_{2G+3G}
\notag \\
=& \frac{- \alpha_s}{2\pi^2 \theta^3 E_N} \int_x^1 dy \int dx_2 \nonumber\\
 &\times \frac{y-x}{y^3} \Biggr \{ \frac{x}{\pi y} T_F (x_2,x_2+y) -\frac{2 x}{y(y-x_2) } 
\bigl[ N(y,x_2) - N(y-x_2,y) + 2 N(y-x_2,-x_2) \bigr] \nonumber\\
& \hspace*{1.2cm}+\frac{1}{x_2^2 (y-x_2) } \biggl [ x_2 (y-x) \bigl[ O(y-x_2,y) - N(y-x_2,y) \bigr] 
\nonumber\\
&\hspace*{3.5cm}+ (y^2 + x x_2 -2xy) \bigl[ N(y,x_2) + O(y,x_2) \bigr ]\nonumber\\
&\hspace*{3.5cm} + y(x_2+2x-y) \bigl[ N(y-x_2,-x_2) + O (y-x_2,-x_2) \bigr] \biggr] \Biggr \}~.
\end{align}

\section{Summary }
\label{sec:summary}
In this paper, we have established a sum rule providing the connections between NEECs and  fracture functions. This suggests that fracture functions can essentially serve as the parent functions of NEECs. This sum rule preserves essential correlations between initial and final states, establishing a one-to-one correspondence between fracture functions and NEECs. We demonstrated that this sum rule is applicable to both the bare and renormalized functions, leading to the conclusion that NEECs and fracture functions adhere to the same evolution kernels. We also explored several extensions of this sum rule, including those related to the $N$-point NEECs and TMD NEECs. This framework of sum rules provides a valuable tool for investigating the properties of NEECs through the analysis of fracture functions.

Using the sum rule, we have advanced the studies of the DIS energy pattern in the TFR from recent developments of SIDIS given in~\cite{Chen:2023wsi,Chen:2024brp}. 
  Through investigations up to twist-3, we derived all eighteen energy-pattern SFs in terms of associated NEECs, incorporating polarization effects of the target and lepton beams. Ten SFs contribute at the twist-2 level, with four of these uniquely sensitive to gluonic NEECs with tensor polarizations. These include the linearly polarized gluon NEEC, $h^{t,g}_{1}$, and three T-odd gluonic NEECs: $h^{t,g}_{1L}$, $h^{t,g}_{1T}$, and $h^{tt,g}_{1T}$. The remaining eight SFs are contributed by twist-3 quark NEECs, manifesting in a compact form at tree level. We also introduce various azimuthal and spin asymmetries to measure these NEECs. Additionally, a comparison with the results in the CFR are presented. 

We have investigated the twist-3 matching of the Sivers-type and worm-gear-type quark NEEC at large $\theta$. These NEECs, T-odd and T-even respectively, are governed by distinct perturbative mechanisms. Using the matching formulas of fracture functions in~\cite{Chen:2021vby}, we express these two NEECs in terms of the twist-3 two-parton and three-parton correlation functions. This analysis provides insights into the transitions of SSA and DSA of the DIS energy pattern between the TFR and the CFR.  

Our comprehensive investigation offers a framework for analyzing energy-weighted observables through hadron production processes in the TFR, paving new avenues for nucleon tomography in forthcoming experiments at facilities like JLab and the EIC.

\acknowledgments 
We thank Feng Yuan, Bowen Xiao and Xiaohui Liu for conversations. The work is supported by National Natural Science Foundation of People’s Republic of China Grants No. 12075299, No. 11821505, No. 11935017 and by the Strategic Priority Research Program of Chinese Academy of Sciences, Grant No.~XDB34000000. K.B. Chen is supported by National Natural Science Foundation of China (Grant No. 12005122), Shandong Province Natural Science Foundation (Grant No. ZR2020QA082), and Youth Innovation Team Program of Higher Education Institutions in Shandong Province (Grant No. 2023KJ126). X.B. Tong is supported by the Research Council of Finland, the Centre of Excellence in Quark Matter and supported under the European Union’s Horizon 2020 research and innovation programme by the European Research Council (ERC, grant agreements No. ERC-2023-101123801 GlueSatLight and No. ERC-2018-ADG-835105 YoctoLHC) and by the STRONG-2020 project (grant agreement No. 824093). The content of this article does not reflect the official opinion of the European Union and responsibility for the information and views expressed therein lies entirely with the authors.

\appendix

\section{Connection between the evolution equations}
\label{sec:evolution}
In this appendix, we demonstrate that the renormalization procedure preserves the sum rule between the NEECs and fracture functions. Furthermore, we show the consistency of their evolution equations.

\subsection{Sum rule for the renormalized functions}
 With loss of generality, we focus on the unpolarized quark NEEC, $f_{1}^a$, and the corresponding fracture function $u_1^{a}$. The sum rule connecting these quantities in their bare form is given by~(see~eq.~(\ref{eq:FrFrelation})):\begin{align}
 f_{1,B}^a(x,\theta)=\sum_h\int_{0}^{1-x} d\xi_h  \frac{\pi\xi_h\bs P_{h\perp}^2 }{\theta^2} u_{1,B}^a(x,\xi_h,\bs P_{h\perp}^2)~\bigg\vert_{|\bs P_{h\perp}| = \frac{\theta \xi_h P^+}{\sqrt{2}}}~.
\label{eq:appsum1}
\end{align} 
Here, the bare functions, $f_{1,B}^a$ and $u_{1,B}^a$, suffer from ultraviolet divergences, which require proper renormalization to subtract these divergences.

To extend this sum rule to the renormalized counterparts, we first recall the multiplicative renormalization of the fracture function~\cite{Berera:1995fj,Grazzini:1997ih,Collins:1997sr}:
\begin{align}
 u_{1,R}^a(x,\xi_h,\bs P_{h\perp}^2,\mu)=\sum_{b}\int^{1-\xi_h}_x \frac{d z}{z} Z_{ab}(x/z,\mu) u_{1,B}^b(z,\xi_h,\bs P_{h\perp}^2)~.
 \label{eq:appZ}
\end{align}
where $\mu$ is the renormalization scale, and $Z_{ab}$ is the renormalization factor. The upper limit in the convolution integral is imposed by the kinematic constraint of the fracture functions.

Following eq.~(\ref{eq:appsum1}), we compute the $\xi_h$-integration on the renormalized fracture functions $u_{1,R}^a$, denoted by:
\begin{align}
{\cal F}(x,\theta,\mu)\equiv\sum_h\int_{0}^{1-x} d\xi_h  \frac{\pi\xi_h\bs P_{h\perp}^2 }{\theta^2} u_{1,R}^a(x,\xi_h,\bs P_{h\perp}^2,\mu)~\bigg\vert_{|\bs P_{h\perp}| = \frac{\theta \xi_h P^+}{\sqrt{2}}}~,
\label{eq:appsum2}
\end{align}
If the sum rule remains valid, the above integration should yield a proper definition of the renormalized NEECs. To illustrate this, we first express ${\cal F}(x,\theta,\mu)$ in terms of the bare fracture functions using eq.~(\ref{eq:appZ}):
\begin{align}
{\cal F}(x,\theta,\mu)=\sum_{b} \sum_h\int_{0}^{1-x} d\xi_h \int^{1-\xi_h}_x \frac{d z}{z} \frac{\pi\xi_h\bs P_{h\perp}^2 }{\theta^2} 
Z_{ab}(x/z,\mu) u_{1,B}^b(z,\xi_h,\bs P_{h\perp}^2)~\bigg\vert_{|\bs P_{h\perp}| = \frac{\theta \xi_h P^+}{\sqrt{2}}}~.
\end{align} 
By changing the order of integration between $d\xi_h$ and $dz$, we obtain:
\begin{align} 
{\cal F}(x,\theta,\mu)=&\sum_{b} \int_x^{1} \frac{d z}{z} 
Z_{ab}(x/z,\mu) \sum_h\int_{0}^{1-z} d\xi_h  \frac{\pi\xi_h\bs P_{h\perp}^2 }{\theta^2}  u_{1,B}^b(z,\xi_h,\bs P_{h\perp}^2)~\bigg\vert_{|\bs P_{h\perp}| = \frac{\theta \xi_h P^+}{\sqrt{2}}}~.
\label{eq:apptem1}
\end{align}
 Using eq.~(\ref{eq:appsum1}), we can identify the $\xi_h$-integral in eq.~(\ref{eq:apptem1}) as the bare NEEC $f_{1,B}^a(z,\theta)$. Meanwhile, the finiteness of ${\cal F}(x,\theta,\mu)$ verifies that this bare NEEC can be renormalized by the same renormalization factor $Z_{ab}$ as the fracture function. Thus, one can define the renormalized NEEC as 
\begin{align}
 f_{1,R}^a(x,\theta,\mu)=\sum_{b}\int^1_x \frac{d z}{z} Z_{ab}(x/z,\mu)f_{1,B}^b(z,\theta)~,
  \label{eq:appZ2}
\end{align}
which satisfies $f_{1,R}^a(x,\theta,\mu)={\cal F}(x,\theta,\mu)$.

It now becomes evident that the renormalized fracture functions and NEECs, defined in eq.~(\ref{eq:appZ}) and eq.~(\ref{eq:appZ2}), obey the following sum rule: \begin{align}
 f_{1,R}^a(x,\theta,\mu)=\sum_h\int_{0}^{1-x} d\xi_h  \frac{\pi\xi_h\bs P_{h\perp}^2 }{\theta^2} u_{1,R}^a(x,\xi_h,\bs P_{h\perp}^2,\mu)~\bigg\vert_{|\bs P_{h\perp}| = \frac{\theta \xi_h P^+}{\sqrt{2}}}~.
\label{eq:appsum2}
\end{align} 
Therefore, we conclude that the sum rule given in eq.~(\ref{eq:appsum1}) is preserved once the NEECs and the fracture functions are renormalized in a consistent scheme. In fact, even if inconsistent schemes are used, the sum rule in eq.~(\ref{eq:appsum2}) would be corrected only by finite terms. Nevertheless, in our calculations in section~\ref{sec:SFfactorization}, we have consistently used the $\overline{\text{MS}}$ scheme for convenience.

\subsection{Consistency of the evolution equations}
A key deduction from eqs.~(\ref{eq:appZ}) and~(\ref{eq:appZ2}) is that the unrenormalized NEECs and their associated fracture functions share the same structures of ultraviolet divergences. Therefore, after renormalization, they follow the same evolution kernel, regardless of the chosen renormalization schemes.

This consistency can also be demonstrated by directly applying the derivative with respect to the renormalization scale $\mu$ in eqs.~(\ref{eq:appZ}) and~(\ref{eq:appZ2}), respectively. First, according to~\cite{Berera:1995fj,Grazzini:1997ih,Chai:2019ykk}, we know that the collinear unpolarized fracture functions $u_{1,R}^{a}$ obey the standard DGLAP evolution equation. Explicitly, taking the derivative of the renormalized fracture function in eq.~(\ref{eq:appZ}) yields: 
\begin{align}
 \frac{d}{d \ln \mu} u_{1,R}^a(x,\xi_h,\bs P_{h\perp}^2,\mu)=\sum_b \int_{x}^{1-\xi_h} \frac{d z}{z} P_{ab}\left(x / z,\alpha_s\right) u_{1,R}^b(z,\xi_h,\bs P_{h\perp}^2,\mu)~,
 \label{eq:appDGLAP}
\end{align} 
where $ P_{ab}$ represents the unpolarized splitting kernels. Meanwhile, the associated renormalization factor evolves according to
\begin{align}
 \frac{d}{d \ln \mu}  Z_{ab}(z,\mu)=\sum_c\int_0^1 \frac{d z'}{z'} P_{ac}(z/z',\alpha_s) Z_{cb}(z',\mu)~.
\end{align}
Given that the renormalized NEECs in eq.~(\ref{eq:appZ2}) utilize the same renormalization factors as the fracture functions, taking their derivative results in:\begin{align}
\frac{d}{d \ln \mu} f_{1,R}^a(x,\theta,\mu) 
=&\sum_b  \int_x^{1} \frac{d z}{z} P_{ab}\left(x / z\right)  f_{1,R}^b(z,\theta,\mu)~.
\end{align}
This derivation explicitly confirms that the evolution of the NEECs mirrors the evolution of the associated fracture functions.

Additionally, one can directly utilize the sum rule in eq.~(\ref{eq:appsum2}) to derive the evolution equations for NEECs from those of fracture functions. This alternative approach also confirms the consistency observed here.

\section{$N$-point NEEC and $N$-hadron fracture functions\label{sec:semiEEC}}  
In this appendix, we present the sum rule between $N$-point NEEC  and $N$-hadron fracture functions.
We begin by introducing the $N$-hadron fracture functions for quarks~\cite{Ceccopieri:2007th}, defined through the following correlation matrix: 
\begin{align}
{\cal M}^{q,N}_{ij,\text{FrF}}& (x,\{\xi_{h_a},\bs P_{h_a\perp}\}) = \int \frac{d\eta^-}{2\pi} e^{-ixP^+\eta^-} \sum_X\int\frac{d^3 \bs P_X}{2 E_X(2\pi)^3}\Big(\prod_a^N\frac{1}{2\xi_{h_a}(2\pi)^{3}}\Big)
\notag \\ 
& \times \langle PS|\bar \psi_j(\eta^-) {\cal L}_n^{\dagger}(\eta^-) |P_{h_1}\cdots P_{h_N} X \rangle \langle X P_{h_1}\cdots P_{h_N}| {\cal L}_n(0) \psi_i(0) |PS\rangle~,
\end{align} 
where $P_{h_a}$ denotes the momentum the detected hadron $h_a$, and $\xi_{h_a}=P_{h_a}^+/P^+$. Here, we include the $P_{h_a\perp}$-dependence.

Next, we consider the $N$-point NEECs~\cite{Liu:2022wop}. The correlation matrix for quarks is given as: 
 \begin{align}
{\cal M}_{ij,{\rm EEC}}^{q,N}&(x,\{\theta_a,\phi_a\}) 
\notag \\ 
=& \int \frac{d\eta^-}{2\pi} e^{-ix P^+ \eta^-} \langle PS|\bar \psi_j(\eta^-) {\cal L}_n^{\dagger}(\eta^-)\Big(\prod_a^N {\cal E}(\theta_a,\phi_a) \Big){\cal L}_n(0) \psi_i(0) |PS\rangle~,
\end{align}
where $\theta_a$ and $\phi_a$ denotes the polar angle and azimuthal angle of the energy flow labeled by $a$.

Following the discussion in section~\ref{sec:connection}, the above two correlation matrices can be connected as follows:
 \begin{align}
{\cal M}_{ij,{\rm EEC}}^{q,N}&(x,\{\theta_a,\phi_a\}) 
\notag \\
=&\sum_{h_1,\cdots,h_N}\int \prod_a^N \Big[d\xi_{h_a} d^2 \bs P_{a\perp}  \delta(\theta^2_a-\theta^2_{h_a})\delta(\phi_a-\phi_{h_a})  
\xi_{h_i} \Big]{\cal M}^{q,N}_{ij,\text{FrF}}(x,\{\xi_{h_a},\bs P_{h_a\perp}\})~.
\label{eq:relationAP1}
\end{align} 
where the sum is over all the combinations of $\{h_1,\cdots,h_N\}$.

Furthermore, one can generalized the above analysis to semi-inclusive energy correlators~\cite{Liu:2024kqt} in the TFR. For example, let us consider the $N$-point energy correlator with an additional hadron $h$ detected in the TFR.
 \begin{align}
{\cal M}_{ij,{\rm EEC}}^{q,N}&(x,\{\theta_a,\phi_a\},\{\xi_h,\bs P_{h\perp}\}) 
=\int \frac{d\eta^-}{2\pi} e^{-ix P^+ \eta^-} \sum_X\int\frac{d^3 \bs P_X}{2 E_X(2\pi)^3}
\notag \\ 
&\times\langle PS|\bar \psi_j(\eta^-) {\cal L}_n^{\dagger}(\eta^-)\Big(\prod_a^N {\cal E}(\theta_a,\phi_a) \Big)|P_{h}X \rangle \langle X P_{h}| {\cal L}_n(0) \psi_i(0) | {\cal L}_n(0) \psi_i(0) |PS\rangle~.
\end{align}
By the same arguments given in section~\ref{sec:connection}, one can show that this semi-inclusive energy correlator can be connected to the $(N+1)$-point energy correlator through the following relation:
 \begin{align}
{\cal M}_{ij,{\rm EEC}}^{q,(N+1)}&(x,\{\theta_a,\phi_a\}) 
\notag \\
=&\sum_{h}\int \xi_h d\xi_hd^2 \bs P_{h\perp} \delta(\theta^2-\theta^2_h)\delta(\phi-\phi_h) 
{\cal M}_{ij,{\rm EEC}}^{q,N}(x,\{\theta_a,\phi_a\},\{\xi_h,\bs P_{h\perp}\}) 
~.
\label{eq:relationAP2}
\end{align} 

\section{TMD quark NEECs and fracture functions}
\label{sec:TMD_NEEC}
We first introduce the TMD quark fracture functions for observing an unpoarlized hadron in a spin-1/2 target. The associated correlation matrix is defined as
\begin{align}
\Phi_{ij,\rm FrF}^q(x, \bs k_\perp,\xi_h, \bs P_{h\perp}) &= \int \frac{d\eta^- d^2 \bs \eta_\perp}{2\xi_h (2\pi)^6} e^{i xP^+ \eta^- - i \bs k_\perp \cdot \bs \eta_\perp} \sum_X \int \frac{d^3 \bs P_X}{(2\pi)^3 2E_X} \nonumber\\
& \times \langle PS|\bar \psi_j(0,\eta^-,\bs\eta_\perp) {\cal L}_n^{\dagger}(0,\eta^-,\bs\eta_\perp) |P_h X \rangle \langle X P_h| {\cal L}_n(0) \psi_i(0) |PS\rangle~.
\end{align}
The leading twist decomposition of this correlation matrix is given by~\cite{Anselmino:2011ss}:
\begin{align}
\Phi_{ij,\rm FrF}^q &= \frac{(\gamma^-)_{ij}}{2N_c}\Biggl( \hat u_1^q - \frac{P_{h\perp} \cdot \tilde S_\perp}{M_h} \hat u_{1T}^{h,q} - \frac{k_{\perp} \cdot \tilde S_\perp}{M} \hat u_{1T}^{\perp,q} + S_L \frac{k_\perp \cdot \tilde P_{h\perp}}{MM_h} \hat u_{1L}^{\perp h,q} \Biggr) \nonumber\\
& + \frac{(\gamma_5 \gamma^-)_{ij}}{2N_c}\Biggl( S_L \hat l_{1L}^q - \frac{P_{h\perp} \cdot S_\perp}{M_h} \hat l_{1T}^{h,q} - \frac{k_{\perp} \cdot S_\perp}{M} \hat l_{1T}^{\perp,q} + \frac{k_\perp \cdot \tilde P_{h\perp}}{MM_h} \hat l_1^{\perp h,q} \Biggr) \nonumber\\
& + \frac{(i\sigma^{\rho-}\gamma_5)_{ij}}{2N_c} \Biggl( S_\perp^\rho \hat t_{1T}^q + S_L \frac{P_{h\perp}^\rho}{M_h} \hat t_{1L}^{h,q} + S_L \frac{k_{\perp}^\rho}{M} \hat t_{1L}^{\perp,q} - \frac{P_{h\perp} \cdot S_\perp}{M_h^2} P_{h\perp}^\rho \hat t_{1T}^{hh,q} \nonumber\\
&\quad - \frac{k_{\perp} \cdot S_\perp}{M^2} k_{\perp}^\rho \hat t_{1T}^{\perp\perp,q} - \frac{(k_\perp \cdot S_\perp) P_{h\perp}^\rho - (k_\perp \cdot S_\perp) k_\perp^\rho}{MM_h} \hat t_{1T}^{\perp h,q} + \frac{\tilde k_{\perp}^\rho}{M} \hat t_1^{\perp,q} + \frac{\tilde P_{h\perp}^\rho}{M_h} \hat t_1^{h,q} \Biggr)~,
\label{eq:TMD-FrFs}
\end{align}
where $M_h$ is the mass of the detected hadron $h$, and $M$ is the target mass.

By using the energy flow operator in eq.~(\ref{eq:Eflow}), one can also define the correlation matrix for TMD quark NEECs:
\begin{align}
\Phi_{ij,\rm EEC}^q(x, \bs k_\perp,\theta, \phi) =& \int \frac{d\eta^- d^2 \bs \eta_\perp}{(2\pi)^3} e^{i xP^+ \eta^- - i \bs k_\perp \cdot \bs \eta_\perp} \nonumber\\
& \times \langle PS|\bar \psi_j(0,\eta^-,\bs\eta_\perp) {\cal L}_n^{\dagger}(0,\eta^-,\bs\eta_\perp) {\cal E}(\theta,\phi) {\cal L}_n(0) \psi_i(0) |PS\rangle~.
\end{align}
The correlation matrix $\Phi_{ij,\rm EEC}^q$ can be decomposed similarly to $\Phi_{ij,\rm FrF}^q$ as
\begin{align}
\Phi_{ij,\rm EEC}^q &= \frac{(\gamma^-)_{ij}}{2N_c}\Biggl( \frac{1}{2\pi} \hat f_1^q - n_t \cdot \tilde S_\perp \hat f_{1T}^{t,q} - \frac{k_{\perp} \cdot \tilde S_\perp}{2\pi M} \hat f_{1T}^{\perp,q} + S_L \frac{k_\perp \cdot \tilde n_t}{M} \hat f_{1L}^{\perp t,q} \Biggr) \nonumber\\
& + \frac{(\gamma_5 \gamma^-)_{ij}}{2N_c}\Biggl( S_L \frac{1}{2\pi} \hat g_{1L}^q - n_t \cdot S_\perp \hat g_{1T}^{t,q} - \frac{k_{\perp} \cdot S_\perp}{2\pi M} \hat g_{1T}^{\perp,q} + \frac{k_\perp \cdot \tilde n_t}{M} \hat g_1^{\perp t,q} \Biggr) \nonumber\\
& + \frac{(i\sigma^{\rho-}\gamma_5)_{ij}}{2N_c} \Biggl( S_\perp^\rho \frac{1}{2\pi} \hat h_{1T}^q + S_L n_t^\rho \hat h_{1L}^{t,q} + S_L \frac{k_{\perp}^\rho}{2\pi M} \hat h_{1L}^{\perp,q} - (n_t \cdot S_\perp) n_t^\rho \hat h_{1T}^{tt,q} \nonumber\\
&\quad - \frac{k_{\perp} \cdot S_\perp}{2\pi M^2} k_{\perp}^\rho \hat h_{1T}^{\perp\perp,q} - \frac{(k_\perp \cdot S_\perp) n_t^\rho - (n_t \cdot S_\perp) k_\perp^\rho}{M} \hat h_{1T}^{\perp t,q} + \frac{\tilde k_{\perp}^\rho}{2\pi M} \hat h_1^{\perp,q} + \tilde n_t^\rho \hat h_1^{t,q} \Biggr) ~.
\label{eq:TMD-NEECs}
\end{align}
The sum rules between the TMD quark NEECs and the TMD quark fracture functions can be derived akin to the collinear case in section~\ref{sec:connection}:
 \begin{align}
\Phi_{ij,\rm EEC}^q(x, \bs k_\perp,\theta, \phi)
=&\sum_{h}\int \xi_h d\xi_hd^2 \bs P_{h\perp} \delta(\theta^2-\theta^2_h)\delta(\phi-\phi_h)
 \Phi_{ij,\rm FrF}^q(x, \bs k_\perp,\xi_h, \bs P_{h\perp})
~.
\end{align} 
 As a result, we have
\begin{align}
& \hat f_1^q (x, \bs k_\perp, \theta) = 2\pi \sumint \hat u_1^q (x, \bs k_\perp, \xi_h, \bs P_{h\perp})~,
 \qquad \hat f_{1T}^{t,q} (x, \bs k_\perp, \theta) = \sumint \frac{|\bs P_{h\perp}|}{M_h} \hat u_{1T}^{h,q} (x, \bs k_\perp, \xi_h, \bs P_{h\perp})~, \nonumber\\
& \hat f_{1T}^{\perp,q} (x, \bs k_\perp, \theta) = 2\pi \sumint \hat u_{1T}^{\perp,q} (x, \bs k_\perp, \xi_h, \bs P_{h\perp})~,
 \quad \hat f_{1L}^{\perp t,q} (x, \bs k_\perp, \theta) = \sumint \frac{|\bs P_{h\perp}|}{M_h} \hat u_{1L}^{\perp h,q} (x, \bs k_\perp, \xi_h, \bs P_{h\perp})~, \nonumber\\
& \hat g_{1L}^q (x, \bs k_\perp, \theta) = 2\pi \sumint \hat l_{1L}^q (x, \bs k_\perp, \xi_h, \bs P_{h\perp})~,
 \qquad \hat g_{1T}^{t,q} (x, \bs k_\perp, \theta) = \sumint \frac{|\bs P_{h\perp}|}{M_h} \hat l_{1T}^{h,q} (x, \bs k_\perp, \xi_h, \bs P_{h\perp})~, \nonumber\\
& \hat g_{1T}^{\perp,q} (x, \bs k_\perp, \theta) = 2\pi \sumint \hat l_{1T}^{\perp,q} (x, \bs k_\perp, \xi_h, \bs P_{h\perp})~,
 \qquad \hat g_1^{\perp t,q} (x, \bs k_\perp, \theta) = \sumint \frac{|\bs P_{h\perp}|}{M_h} \hat l_1^{\perp h,q} (x, \bs k_\perp, \xi_h, \bs P_{h\perp})~, \nonumber\\
& \hat h_{1T}^q (x, \bs k_\perp, \theta) = 2\pi \sumint \hat t_{1T}^q (x, \bs k_\perp, \xi_h, \bs P_{h\perp}), \qquad \hat h_{1L}^{t,q} (x, \bs k_\perp, \theta) = \sumint \frac{|\bs P_{h\perp}|}{M_h} \hat t_{1L}^{h,q} (x, \bs k_\perp, \xi_h, \bs P_{h\perp})~, \nonumber\\
& \hat h_{1L}^{\perp,q} (x, \bs k_\perp, \theta) = 2\pi \sumint \hat t_{1L}^{\perp,q} (x, \bs k_\perp, \xi_h, \bs P_{h\perp})~,
 \qquad \hat h_{1T}^{tt,q} (x, \bs k_\perp, \theta) = \sumint \frac{\bs P_{h\perp}^2}{M_h^2} \hat t_{1T}^{hh,q} (x, \bs k_\perp, \xi_h, \bs P_{h\perp})~, \nonumber\\
& \hat h_{1T}^{\perp\perp,q} (x, \bs k_\perp, \theta) = 2\pi \sumint \hat t_{1T}^{\perp\perp,q} (x, \bs k_\perp, \xi_h, \bs P_{h\perp})~,
 \quad \hat h_{1T}^{\perp t,q} (x, \bs k_\perp, \theta) = \sumint \frac{|\bs P_{h\perp}|}{M_h} \hat t_{1T}^{\perp h,q} (x, \bs k_\perp, \xi_h, \bs P_{h\perp})~, \nonumber\\
& \hat h_1^{\perp,q} (x, \bs k_\perp, \theta) = 2\pi \sumint \hat t_1^{\perp,q} (x, \bs k_\perp, \xi_h, \bs P_{h\perp})~, \qquad \hat h_1^{t,q} (x, \bs k_\perp, \theta) = \sumint \frac{|\bs P_{h\perp}|}{M_h} \hat t_1^{h,q} (x, \bs k_\perp, \xi_h, \bs P_{h\perp})~,
\label{eq:Relation-TMD-NEEC-FrF}
\end{align}
where we have used the notation $\sumint$ defined in eq.~(\ref{eq:sumint}).

\section{Twist-3 parton distributions}
\label{sec:twist3PDF}
A complete set of independent twist-3 parton distributions $(q_T, q_\partial,T_F,T_\Delta, O,N)$ has been employed in section~\ref{sec:matching} to study the matching of quark NEECs. In this appendix, we provide definitions of these distributions. For detailed discussions, we refer to ref.~\cite{Chen:2021vby} and the references therein.

 For two-parton correlations, we introduce the following twist-3 quark distributions~\cite{Chen:2015uqa}: 
\begin{align} 
  q_T (x) S_\perp^\mu  =&  P^+ \int \frac{d\lambda}{ 4\pi } e^{- i x\lambda  P^+} \langle PS  \vert \bar \psi(\lambda n) {\mathcal L}^\dagger_n(\lambda n)  \gamma_\perp^\mu \gamma_5 
  {\mathcal L}_n(0) \psi(0) \vert PS  \rangle~,
\nonumber\\  
  -i q_\partial (x) S_\perp^\mu =&    \int \frac{d\lambda}{ 4\pi } e^{- i x\lambda  P^+} \langle PS  \vert \bar \psi(\lambda n) {\mathcal L}^\dagger_n(\lambda n)  \gamma^+ \gamma_5 \partial_\perp^{\mu}   
  \left ( {\mathcal L}_n  \psi \right ) (0) \vert PS \rangle~.
\label{tw31}  
\end{align}
For quark-gluon-quark correlations, we use the $F$-type twist-3 distributions:
\begin{align}
T_F (x_1,x_2) \tilde S^\mu_\perp=&  g_s \int\frac{d y_1 dy_2}{4\pi} e^{-iy_1x_1 P^+ -i y_2 (x_2-x_1) P^+} 
   \langle  PS \vert \bar \psi (y_1 n) \gamma^+  G^{+\mu}(y_2 n)  \psi (0) \vert  PS \rangle~,
   \notag \\ 
   T_\Delta (x_1,x_2) i S^\mu_\perp=& g_s \int\frac{d y_1 dy_2}{4\pi} e^{-iy_1x_1 P^+ -i y_2 (x_2-x_1) P^+} 
   \langle  PS \vert \bar \psi (y_1 n) \gamma^+ \gamma_5  G^{+\mu}(y_2 n)  \psi (0) \vert  PS \rangle~,
\end{align}
where the gauge links are implied for short notations. These distributions satisfy the following symmetries~\cite{Efremov:1981sh,Efremov:1984ip,Qiu:1991pp,Qiu:1991wg,Qiu:1998ia,Kodaira:1998jn,Zhou:2009jm}:
\begin{align} 
     T_F (x_1,x_2) =  T_F(x_2,x_1)~, \quad \quad  T_{\Delta} (x_1,x_2) = - T_{\Delta} (x_2,x_1)~.             
\label{TW3P}
\end{align} 
For three-gluon correlations, we use the twist-3 gluon distributions $O$ and $N$ defined from the following matrix:
\begin{align} 
& \frac{i^3 g_s}{P^+}  \int \frac{d\lambda_1}{2\pi }\frac{d\lambda_2}{2\pi} 
   e^{i\lambda_1 x_1 P^+ + i\lambda_2 (x_2-x_1)P^+} \langle PS  \vert G^{a,+\alpha} 
   (\lambda_1 n)  G^{c,+\gamma}(\lambda_2 n) G^{b,+\beta} (0) \vert PS  \rangle 
\nonumber\\
   = & \frac{N_c}{(N_c^2-1)(N_c^2-4)} d^{abc} O^{\alpha\beta\gamma}(x_1,x_2) - \frac{i }{N_c(N_c^2-1) } f^{abc} N^{\alpha\beta\gamma}(x_1,x_2)~, 
\label{NO3G}    
\end{align} 
where all indices $\alpha,\beta$ and $\gamma$ are transverse, and the two tensors take the form~\cite{Ji:1992eu,Beppu:2010qn,Koike:2011ns}:
\begin{align}
O^{\alpha\beta\gamma}(x_1,x_2) =& -2 i \biggr [ O(x_1,x_2) g^{\alpha\beta}_\perp \tilde S_\perp^\gamma + 
  O(x_2,x_2-x_1) g^{\beta\gamma}_\perp \tilde S_\perp^\alpha + O(x_1,x_1-x_2) g^{\gamma\alpha}_\perp \tilde S_\perp^\beta \biggr ]~, 
\nonumber\\
N^{\alpha\beta\gamma}(x_1,x_2) =& -2 i \biggr [ N(x_1,x_2) g^{\alpha\beta}_\perp \tilde S_\perp^\gamma - 
  N(x_2,x_2-x_1) g^{\beta\gamma}_\perp \tilde S_\perp^\alpha - N(x_1,x_1-x_2) g^{\gamma\alpha}_\perp \tilde S_\perp^\beta \biggr ]~.
\label{NOG3}           
\end{align}
The functions $O$ and $N$ obey the following properties:
\begin{align} 
    & O(x_1,x_2) =O(x_2,x_1)~,\quad O(x_1,x_2) = O(-x_1,-x_2)~, 
\nonumber\\ 
   & N(x_1,x_2) = N(x_2,x_1)~, \quad
N(x_1,x_2) = -N(-x_1,-x_2)~. 
\end{align}           
In our conventions, all twist-3 parton distributions have the dimension $1$ in mass and are proportional 
to $\Lambda_{\text{QCD}}$.

\bibliographystyle{JHEP}
\bibliography{Paper_JHEP.bib}
\end{document}